\documentclass[aps,prl,preprint,superscriptaddress]{revtex4-1}

\newcommand{\tr}{\rm tr \,}
\usepackage{graphicx}
\usepackage{amsmath}
\usepackage{amssymb}
\usepackage{slashed}

\usepackage{hyperref}
\usepackage{color}

\begin{document}


\title{On the chiral expansion of vector meson masses}


\author{R. Bavontaweepanya}
\affiliation{Mahidol University, Bangkok 10400,  Thailand}
\author{Xiao-Yu Guo}
\affiliation{GSI Helmholtzzentrum f\"ur Schwerionenforschung GmbH, \\Planckstra\ss e 1, 64291 Darmstadt, Germany}
\author{M.F.M. Lutz}
\affiliation{GSI Helmholtzzentrum f\"ur Schwerionenforschung GmbH, \\Planckstra\ss e 1, 64291 Darmstadt, Germany}
\affiliation{Technische Universit\"at Darmstadt, D-64289 Darmstadt, Germany}

\date{\today}

\begin{abstract}
We study the chiral expansion of meson masses and decay constants using a chiral Lagrangian 
that was constructed previously based on the hadrogenesis conjecture. The 
one-loop self energies of the Goldstone bosons and vector mesons are evaluated. It is illustrated that a renormalizeable effective field theory arises once 
specific conditions on the low-energy constants are imposed. For the case where 
the hadrogenesis mass gap scale $\Lambda_{\rm HG}$ is substantially larger than the chiral symmetry breaking scale $\Lambda_\chi$ 
a partial summation scheme is required. All terms proportional to $(M/\Lambda_\chi)^n$ can be summed by a suitable renormalization, where $M$ is the chiral and large-$N_c$ limit 
of the vector meson masses in QCD. The size of loop effects from vector meson degrees of freedom
is illustrated for physical quarks masses. Naturally sized effects are observed that have significant impact on the chiral structure of low-energy QCD with three light flavours. 
\end{abstract}


\maketitle

\section{Introduction}
\label{sec:intro}

Vector meson degrees of freedom are known to play an important role in hadron physics. Since the seminal work of Sakurai \cite{sakurai-book}
pioneering the vector-meson dominance phenomenology there is the quest how such a picture can be related to the underlying fundamental 
theory of strong interactions. To the best knowledge of the authors such a link to  QCD has not been established so far 
{\color{green}\cite{Meissner:1987ge,Bijnens:1995ii,Birse:1996hd,Bijnens:1999sh,Klingl:1996by,Lutz:2001mi,Djukanovic:2004mm,Bruns:2004tj,Djukanovic:2009zn,Lutz:2008km,Terschlusen:2012xw}}. 

There is a rather successful chiral Lagrangian originally constructed by Bando and coauthors, where the vector mesons are considered as non-abelian gauge bosons 
properly coupled to the Goldstone bosons in compliance with the chiral Ward identities of QCD 
{\color{green}\cite{Bando:1987br,Vaffa1984,Harada:2003jx}}. The challenge of such a path is the question whether the effective Lagrangian 
is general enough as to guarantee a systematic link to QCD as its low-energy effective field theory. Is there any power-counting principal that generalizes the 
Lagrangian and permits a consistent renormalization program? An alternative starting point is a chiral Lagrangian where vector meson degrees of freedom are considered as heavy fields initially
{\color{green}\cite{Jenkins:1995vb,Bijnens:1996nq,Bijnens:1997ni,Bijnens:1998di,Djukanovic:2010tb,Bruns:2013tja}}. 
An infinite tower of interaction terms can readily be written down. However, it is unclear how to order this plethora of terms and how to consider the loop effects implied
{\color{green}\cite{Bruns:2004tj,Fuchs:2003sh}}. 
In both approaches the challenge  is caused by meson resonances that are close to the vector mesons in mass
{\color{green}\cite{Kampf:2009jh,Pich:2008jm,Guo:2014yva}}. Is there any rational to construct a chiral Lagrangian 
with vector mesons but leaving out for instance scalar and axial vector mesons? Indeed the hadrogenesis conjecture  proposes such a scenario: meson resonances that are not 
considered as explicit degrees of freedom in the effective Lagrangian may be dynamically generated by coupled-channel dynamics based on that Lagrangian \cite{Lutz:2001yb,Lutz:2001dr,Lutz:2003fm,Kolomeitsev:2003ac,Kolomeitsev:2003kt,Lutz:2007sk}. While for scalar mesons 
such a mechanism is known since the early days of the quark model \cite{vanBeveren:1986ea,Weinstein:1990gu}, only a decade ago one of the authors illustrated  that chiral symmetry predicts 
a spectrum of axial-vector mesons as a consequence of the coupled-channel interactions of the Goldstone bosons with vector mesons \cite{Lutz:2003fm,Kolomeitsev:2003ac,Roca:2005nm,Lutz:2008km,Wagner:2008gz}. While such results 
support the hadrogenesis conjecture it is still an open challenge how to systematize such an approach. 

Recently a possible direct link of the hadrogenesis conjecture to QCD was suggested in \cite{Terschlusen:2012xw}. If chiral QCD at vanishing up, down and strange quark masses is considered in the limit of a 
large number of colors ($N_c$), an infinite tower of discrete states appears \cite{Witten:1979vv}.  While it is established that such a tower of states exists, it is not known from first principal where the levels 
are located. There is no stringent reason that this spectrum resembles closely the excitation spectrum of QCD at finite quark masses and a finite number of colors $N_c$. 
Suppose that there would be a significant mass gap below some hard scale $\Lambda_{HG}$ in chiral QCD at large $N_c$. This would  permit the construction of an effective field theory description for the 
physics below that heavy scale $\Lambda_{HG}$. The relevant degrees of freedom are identified with the states in the spectrum that are below that scale. A possible minimal scenario would be that those relevant 
degrees of freedom are the Goldstone bosons accompanied by the light vector mesons only. Based on this assumption the leading order chiral Lagrangian was constructed in \cite{Terschlusen:2012xw}. The 
power counting is based on the assumption that $M_V/\Lambda_{HG}\sim Q$ is sufficiently small as to arrive at a convergent expansion.

While in \cite{Terschlusen:2012xw} a chiral Lagrangian was constructed according to a dimensional power counting scheme in the presence of a conjectured hadrogenesis scale $\Lambda_{HG}$, its consistency as 
an effective field theory remained an open issue. In particular can it be renormalized convincingly \cite{Terschlusen:2016cfw,Terschlusen:2016kje}? This question will be studied at the one-loop level in this work. 
It is well known from various quantum field theories that the quest of renormalizeability may impose stringent conditions on the form of the effective Lagrangian. 

The work is organized as follows. In section 2 we recall the chiral Lagrangian as constructed in \cite{Terschlusen:2012xw}. Additional terms of order $Q^4 $ are constructed that are required 
for the renormalization of the one-loop contributions to the meson masses. It follows sections 3 and 4 where the one-loop contributions to the  meson masses are computed  and analyzed. 
Explicit results on the scale dependence of the low-energy constants are derived. The importance of explicit vector meson degrees of freedom is illustrated at hand of a series of figures that 
detail the one-loop contributions to the meson masses. In section 5 the decay constants of the Goldstone bosons are considered. Section 6 gives a short summary and outlook.

\newpage 

\section{The chiral Lagrangian with light vector-meson fields}
\label{sec:chiral-lagrangian}

We recall the hadrogenesis Lagrangian as introduced in \cite{Terschlusen:2012xw}. 
A chiral $SU(3)$ Lagrangian is readily constructed utilizing appropriate building blocks 
\cite{Krause:1990,Ecker:1988te,Ecker:1989yg,Birse:1996hd,HerreraSiklody:1996pm,Kaiser:2000gs}. The basic
elements are
\begin{eqnarray}
  U_\mu & = & {\textstyle \frac{1}{2}}\,e^{-i\,\frac{\Phi}{2\,f}} \left(
    \partial_\mu \,e^{i\,\frac{\Phi}{f}} \right) e^{-i\,\frac{\Phi}{2\,f}}
  -{\textstyle \frac{i}{2}}\,e^{-i\,\frac{\Phi}{2\,f}} \,r_\mu\, e^{+i\,\frac{\Phi}{2\,f}}  
  \nonumber \\  && {}
  +{\textstyle \frac{i}{2}}\,e^{+i\,\frac{\Phi}{2\,f}} \,l_\mu\, e^{-i\,\frac{\Phi}{2\,f}} \,,
  \qquad \qquad  \Phi_{\mu \nu} \;, \qquad \qquad H = \frac{1}{\sqrt{6}\,f}\,\tr \,\Phi  \,,
\nonumber\\
  f_{\mu \nu}^{\pm} & = & {\textstyle \frac{1}{2}} \, e^{+i\,\frac{\Phi}{2\,f}}
  \left( \partial_\mu \,l_\nu- \partial_\nu \,l_\nu -i\,[l_\mu ,\,l_\nu]_-\right)
  e^{-i\,\frac{\Phi}{2\,f}}
  \nonumber\\  && {}
  \pm  {\textstyle \frac{1}{2}} \, e^{-i\,\frac{\Phi}{2\,f}}
  \left( \partial_\mu \,r_\nu\,- \partial_\nu \,r_\nu \,-i\,[r_\mu \,,\,r_\nu\,]_-\right)
  e^{+i\,\frac{\Phi}{2\,f}} \,,   
    \label{def-fields}
\end{eqnarray}
where we include a nonet of  pseudoscalar-meson fields
$\Phi(J^P\!\!=\!0^-)$ and a nonet of vector-meson fields  in the antisymmetric tensor representation
$\Phi_{\mu \nu} (J^P\!\!=\!1^-)$. The notations and conventions of \cite{Terschlusen:2012xw} are used through out this work. 
The classical source functions $r_\mu$ and $l_\mu$ in (\ref{def-fields}) are
linear combinations of the vector and axial-vector sources of QCD with $r_\mu = v_\mu+a_\mu$ and $l_\mu = v_\mu-a_\mu$.
Explicit chiral symmetry-breaking effects are included in terms
of scalar and pseudoscalar source fields $\chi_\pm $ proportional to the quark-mass
matrix of QCD
\begin{eqnarray}
\chi_\pm = {\textstyle \frac{1}{2}} \left(
e^{+i\,\frac{\Phi}{2\,f}} \,\chi_0 \,e^{+i\,\frac{\Phi}{2\,f}}
\pm e^{-i\,\frac{\Phi}{2\,f}} \,\chi_0 \,e^{-i\,\frac{\Phi}{2\,f}}
\right) \,,
\label{def-chi}
\end{eqnarray}
where $\chi_0 =2\,B_0\, {\rm diag} (m_u,m_d,m_s)$. 

The covariant derivative $D_\mu$ discriminates flavour octet from flavour singlet fields.  
It is identical for all matrix fields in (\ref{def-fields}) and (\ref{def-chi})
\begin{eqnarray}
&& D_\mu O  =  \partial_\mu O + \big[\Gamma_\mu,\,O \big]_- \,, \qquad  \qquad \qquad 
D_\mu H = \partial_\mu  H  - {\textstyle{ \sqrt{\frac{2}{3}} }}\,\tr (a_\mu) \,,
\nonumber\\
&& \Gamma_\mu  = {\textstyle \frac{1}{2}}\,e^{-i\,\frac{\Phi}{2\,f}} \,
\Big[\partial_\mu -i\,(v_\mu + a_\mu) \Big] \,e^{+i\,\frac{\Phi}{2\,f}}
+{\textstyle \frac{1}{2}}\, e^{+i\,\frac{\Phi}{2\,f}} \,
\Big[\partial_\mu - i\,(v_\mu - a_\mu)\Big] \,e^{-i\,\frac{\Phi}{2\,f}}\,,
\label{def-covariant-derivative}
\end{eqnarray}
with $O\in \{ U_\mu,  \Phi_{\mu \nu}, f^\pm_{\mu\nu},\chi_\pm \}$ and the chiral connection $\Gamma_\mu$. 
In a covariant derivative on the singlet field $H$ the axial source function $a_\mu$ is probed only.

In the following we focus on terms established previously in \cite{Lutz:2008km,Terschlusen:2012xw}
that do not involve the flavour singlet field $H$. At second order the various terms can be grouped into three classes
\begin{eqnarray}
&& {\mathcal L}^{(2)}_{2} = -\frac{1}{4}\,{\tr }\, \Big\{(D^\mu\,\Phi_{\mu \alpha})\,(D_\nu \,\Phi^{\nu \alpha})\Big\}+
\frac{1}{8}\,M^2\,{\tr } \,\Big\{ \Phi^{\mu \nu}\,\Phi_{\mu \nu}\Big\}
\nonumber\\
&& \qquad  
+\, \frac{1}{2}\,f_V\,{\rm tr} \Big\{\Phi^{\mu\nu}\,f^+_{\mu \nu}\Big\}  - f^2\,{\tr}  \big\{ U_\mu \,U^\mu \big\}
  + \frac{1}{2}\,f^2\,{\tr}  \big\{\,\chi_+  \big\}  \,,
\label{def-L22}\\ \nonumber\\
&& {\mathcal L}^{(3)}_2  =  \frac{i}{2}\,f\, h_1\,{\rm tr}\,\Big\{U_\mu\,\Phi^{\mu\nu}\,U_\nu\Big\}
   +\ \frac{i}{8} \,h_2 \,\varepsilon^{\mu\nu\alpha\beta}\,
  {\rm tr} \,\Big\{ \big[\Phi_{\mu \nu},\,(D^\tau \Phi_{\tau \alpha})\big]_+ \,U_\beta\Big\}
\nonumber \\
&&\qquad   -\, \frac{i}{4} \,\frac{M^2}{f} \, h_3 \,{\rm tr}\,\Big\{
  \Phi_{\mu \tau}\,\Phi^{\mu \nu}\,\Phi^{\tau}_{\;\;\, \nu} \Big\}\,,
\label{def-L23}
\\ \nonumber\\
&& {\mathcal L}^{(4)}_2  = 
    \frac{1}{8} \, g_1 \,
    {\tr } \,\Big\{\big[ \Phi_{\mu \nu }\,,U_\alpha \big]_+ \, \big[U^\alpha, \Phi^{\mu \nu} \big]_+\Big\}
    +\frac{1}{8} \, g_2 \,
    {\tr } \,\Big\{\big[ \Phi_{\mu \nu }\,,U_\alpha \big]_- \, \big[U^\alpha, \Phi^{\mu \nu}\big]_- \Big\}
\nonumber \\
&& \qquad
    + \frac{1}{8} \, g_3 \,
    {\tr } \,\Big\{\big[\,U_\mu\,,U^\nu \big]_+ \, \big[\Phi_{\nu \tau}\,, \Phi^{\mu \tau}\big]_+  \Big\}
       +\frac{1}{8} \, g_4 \,
    {\tr } \,\Big\{\big[\,U_\mu\,,U^\nu \big]_- \, \big[\Phi_{\nu \tau} \,, \Phi^{\mu \tau}\big]_- \Big\}
\nonumber \\
&& \qquad + \frac{1}{8} \, g_5 \,
    {\tr } \,\Big\{\big[\Phi^{\mu \tau} , U_\mu\big]_- \, \big[\Phi_{\nu \tau} \,, U^\nu\big]_- \Big\}
     + \frac18 \, \frac{M^2}{f^2} \, g_6 \,
    {\tr } \,\Big\{\big[\Phi_{\mu \nu} \,, \Phi_{\alpha \beta} \big]_+ \,
    \big[\Phi^{\alpha \beta} , \Phi^{\mu \nu}\big]_+ \Big\}
\nonumber \\
&& \qquad
    + \frac18 \, \frac{M^2}{f^2} \, g_7 \,
    {\tr } \,\Big\{\big[\Phi_{\alpha \beta} \,, \Phi_{\mu \nu} \big]_- \,
    \big[\Phi^{\alpha \beta} ,\Phi^{\mu \nu}\big]_- \Big\}
     + \frac18 \, \frac{M^2}{f^2} \, g_8 \,
    {\tr } \,\Big\{ \big[\Phi^{\mu \nu} ,\Phi_{\mu \beta} \big]_+ \,
    \big[\Phi_{\alpha \nu} \,, \Phi^{\alpha \beta}\big]_+ \Big\}
\nonumber \\
&& \qquad
    + \frac18 \, \frac{M^2}{f^2} \, g_9 \,
    {\tr } \,\Big\{ \big[\Phi^{\mu \nu} ,\Phi_{\mu \beta} \big]_- \,
    \big[\Phi_{\alpha \nu} \,, \Phi^{\alpha \beta}\big]_- \Big\}\,,
\label{def-L24}
\end{eqnarray}
where we recall the counting scheme with $D_\mu , U_\mu\sim Q$ but $ \chi_\pm \sim Q^2$. The scale $M\sim Q$ is counted as order one, if 
it is probed relative to the hadrogenesis scale $\Lambda_{HG}$ with $M/\Lambda_{HG} \sim Q$ (see \cite{Terschlusen:2012xw}).  

It is well known from various quantum field theories that the quest of renormalizeability may impose stringent conditions on the form of the effective Lagrangian.  Indeed 
we already omitted three terms initially suggested in \cite{Terschlusen:2012xw} to enter the Lagrangian at oder $Q^2$. We anticipate the outcome of our study which requires 
that these three terms contribute at oder $Q^4$ only. The first term 
\begin{eqnarray}
 \frac{1}{8}\,b_D \,{\tr } \,\Big\{ \Phi^{\mu \nu}\,\Phi_{\mu \nu}\,\chi_+\Big\} \to \,\frac{1}{8}\,b_1 \,M^2\,{\tr } \,\Big\{ \Phi^{\mu \nu}\,\Phi_{\mu \nu}\,\chi_+\Big\}\,,
\end{eqnarray}
breaks chiral symmetry explicitly. By assigning to its structure the factor $M^2$ it is moved from ${\mathcal L}_{2}$ to ${\mathcal L}_{4}$. This implies that all vector meson masses 
are given by $M$ at leading order in our counting scheme. Without such a property we do not see any path for a consistent renormalization program. The case for the other 
two terms 
\begin{eqnarray}
&& \qquad 
   \frac{i}{8} \, h_4 \, \varepsilon^{\mu \nu \alpha \beta}\,
 { \tr} \Big\{ \big[ (D_\alpha \Phi_{\mu \nu}), \, \Phi_{\tau \beta}\big]_+ \, U^\tau\Big\}
 \,, \qquad  \,\frac{i}{4} \,h_5 \,\varepsilon^{\mu\nu\alpha\beta} \,
  {\rm tr} \,\Big\{ \Phi_{\mu\nu}\, \chi_-\, \Phi_{\alpha \beta} \Big\}\,,
\end{eqnarray}
is more intricate. If considered at order $Q^2 $ they would generate a scale-dependence at the one-loop level that cannot be absorbed into 
the available counter terms at oder $Q^4$. Therefore we insist on $h_{4,5} \to M^2\,h_{4,5} $ as well. Thus such terms contribute to ${\mathcal L}_{5}$ and therefore 
turn irrelevant for the one-loop study of this work.

Some of the low-energy 
parameters have been estimated before in \cite{Lutz:2008km,Terschlusen:2012xw} with
\begin{eqnarray}
&&h_1 = f_V \,h_P/f \simeq 2.5 \pm 0.25\,, \qquad \qquad h_2 = h_A = 2.33\pm 0.03\,,\qquad 
\nonumber\\
&& h_3 = f\,h_V/f_V \simeq 0.05 \,,
\label{parameter-set-I}
\end{eqnarray}
where $h_P,h_A, h_V$ are the low-energy constants as introduced in  \cite{Terschlusen:2012xw}. The parameter $f \simeq 90$ MeV is the chiral limit value 
of the pion or kaon decay constant. The tree-level estimate for the parameter $b_1$ from \cite{Terschlusen:2012xw} should be rejected since 
according to our findings it should be determined in the presence of one-loop effects. 
For the remaining constants $g_i$ so far no reliable estimate exists.

Since we will compute the one-loop contributions to the Goldstone boson and vector meson self energies we need to collect an appropriate 
set of counter terms to renormalize their scale dependent parts. According to our power counting the latter are expected to be of order four.  
While the Goldstone boson sector  \cite{Gasser:1984gg}  is well established 
\begin{eqnarray}
&& \mathcal{L}^{(P)}_4  =  16\,L_1\,({\tr}\{U_\mu U^\mu\})^2 + 16\,L_2\,{\tr} \{U_\mu U_\nu\}{\tr}\{U^\mu U^\nu\} 
+ 16\,L_3\,{\tr}\{U_\mu U^\mu U_\nu U^\nu\} 
\nonumber\\
&& \qquad \, - \,8\,L_4\,{\tr}\{U_\mu U^\mu\}\,{\tr}\{\chi_+\}- 8\,L_5\,{\tr}\{U_\mu U^\mu \chi_+\} 
+ 4\,L_6\,{\tr}\{\chi_+\}\,{\tr}\{\chi_+\} 
\nonumber\\
&& \qquad \,+\, \,4\,L_7\,{\tr}\{\chi_-\}\,{\tr}\{\chi_-\} 
+2\,L_8\,{\tr}\{\chi_+ \chi_+ + \chi_- \chi_- \}\,,
\label{def-L42}
\end{eqnarray}
this is not the case for the terms involving the light vector mesons. A complete construction of the complete fourth order Lagrangian in the 
presence of vector meson fields is beyond the scope of our work. Here we focus on the terms which involve two vector meson fields and at most two $\chi_+$ fields. All together we have 
\begin{eqnarray}
&& \mathcal{L}_{4}^{(V)} =  \frac{e_1}{8}\,M^4\,{\tr } \,\Big\{ \Phi^{\mu \nu}\,\Phi_{\mu \nu}\Big\} + 
\frac{e_2}{8}\,M^4\,{\tr\{\Phi_{\mu\nu}\}\,\tr\{\Phi^{\mu\nu}\} }
\nonumber\\
&& \qquad + \, \frac{b_1}{8}\, M^2\,{\tr } \,\Big\{ \Phi^{\mu \nu}\,\Phi_{\mu \nu}\,\chi_+\Big\} 
+ \frac{b_2}{8}\,M^2 \,{\tr\{\Phi_{\mu\nu}\,\Phi^{\mu\nu}\}\,\tr\{\chi_+\} }+  \frac{b_3}{8}\,M^2\,{\tr\{\Phi_{\mu\nu}\}\,\tr\{\Phi^{\mu\nu}\,\chi_+\}}
\nonumber\\ 
&& \qquad + \, \frac{c_1}{8}\,{\tr\,\{\Phi_{\mu\nu}\,\chi_+\,\Phi^{\mu \nu}\,\chi_+\}} +  \frac{c_2}{8}\,\tr\,\{\Phi_{\mu\nu}\,\Phi^{\mu\nu}\,\chi_+^2\} \,,
\label{counter-terms}
\end{eqnarray}
where we do not consider corresponding terms with further number of traces at this order. 
Any term that involves additional flavour traces are suppressed by the factor $1/N_c$ at least. In our scheme this is translated into the  factor  $M^2$ such as to 
transport this suppression factor into the dimensional counting rule.  
We further illustrate our construction principal by a partial list of $Q^6$ and $Q^8$ terms
\begin{eqnarray}
&& \mathcal{L}_{6}^{(V)} = \frac{c_3}{8}\,M^2\,{\tr\,\{\Phi_{\mu\nu}\,\Phi^{\mu\nu}\}\,\tr\,\{\chi_+^2\}} + \frac{c_4}{8}\,M^2\,\tr\,\{\Phi_{\mu\nu}\,\Phi^{\mu\nu}\,\chi_+\}\,\tr\,\{\chi_+\} 
\nonumber\\
&& \qquad  + \,   \frac{c_5}{8}\,M^2\,\tr\,\{\Phi^{\mu\nu}\,\chi_+\}\,\tr\,\{\Phi_{\mu\nu}\chi_+\} +\frac{c_6}{8}\,M^2\,\tr\,\{\Phi^{\mu\nu}\}\,\tr\,\{\Phi_{\mu\nu}\,\chi_+^2\}
\nonumber\\
&& \qquad +\, \frac{b_4}{8}\,M^4\,{\tr\{\Phi_{\mu\nu}\}\,\tr\{\Phi^{\mu\nu}\}\,\tr\,\{\chi_+\}}  \,,
\nonumber\\
&& \mathcal{L}_{8}^{(V)} =  \frac{c_7}{8}\,M^4\,{\tr\,\{\Phi^{\mu\nu}\,\Phi_{\mu\nu}\}\,\tr\,\{\chi_+\}\,\tr\,\{\chi_+\}}
+\frac{c_8}{8}\,M^4\,{\tr\,\{\Phi^{\mu\nu}\}\,\tr\,\{\Phi_{\mu\nu}\,\chi_+\}\,\tr\,\{\chi_+\} }
\nonumber\\
&& \qquad  + \,\frac{c_9}{8}\,M^4\,\tr\,\{\Phi^{\mu\nu}\}\,\tr\,\{\Phi_{\mu\nu}\}\,\tr\,\{\chi_+^2\}\,.
\label{def-extra-terms}
\end{eqnarray}
Note that a further term with four traces is redundant as it can be generated by a suitable combination of terms presented in (\ref{def-extra-terms}). 

For none of the dimension full parameters $c_i$ a numerical estimate is available. According to the hadrogenesis conjecture we expect for instance 
$c^{\rm ren }_{1,2} \sim \Lambda^{-2}_{HG}$ with $\Lambda_{HG}> 2$ GeV for suitably renormalized low-energy parameters.

\clearpage

\section{Vector meson masses at the one-loop level}
\label{sec:vector-mesons}

We begin with a collection of all tree-level 
expressions for the vector meson masses from $\mathcal L_4^{(V)}$ in (\ref{counter-terms}) 
for which we find the result
\begin{eqnarray}
&& \Pi^{\rm tree}_{\rho}  =   e_1\,M^4 + 2\,b_1\,M^2\,B_0\,m + 2\,b_2\,M^2\,B_0\,(2\,m+m_s)+ 4\,(c_1+c_2)\,B_0^2\,m^2 \,,
\nonumber\\
&&  \Pi^{\rm tree}_{\omega} =  e_1\,M^4 + 2\,e_2\,M^4 + 2\,(b_1+2\,b_3)\,M^2\,B_0\,m + 2\,b_2\,M^2\,B_0\,(2\,m+m_s)  
\nonumber\\
&& \qquad +\, 4\,(c_1+c_2 )\,B_0^2\,m^2 \,,
\nonumber\\
&&  \Pi^{\rm tree}_{K^*}  =  e_1\,M^4+ b_1\,M^2\,B_0\,(m+m_s) + 2\,b_2\,M^2\,B_0\,(2\,m+m_s) 
\nonumber\\
&& \qquad +\, 4\,(c_1+c_2)\,B_0^2\,m\,m_s
+ 2\,c_2\,B_0^2\,(m-m_s)^2 \,,
\nonumber\\
&&  \Pi^{\rm tree}_{\phi} =   (e_1+e_2)\,M^4 + 2\,(b_1+b_3)\,M^2\,B_0\,m_s + 2\,b_2\,M^2\,B_0\,(2\,m+m_s) 
\nonumber\\
&& \qquad   +\, 4\,( c_1+c_2 )\,B_0^2\,m_s^2 \,,
\nonumber\\
&&  \Pi^{\rm tree}_{\omega \phi} = \sqrt{2}\,e_2\,M^4 + \sqrt{2}\,b_3\,M^2\,B_0\,(m+m_s) \,  ,
\label{tree-level-V}
 \end{eqnarray}
where  a projection of the polarization tensor on its mass component is understood. One may introduce an $\omega-\phi$ mixing angle $\epsilon$ by 
\cite{Okubo:1963fa,Kucukarslan:2006wk}
\begin{eqnarray}
&& \omega =\omega' \, \cos \epsilon +\phi' \,\sin \epsilon \,, \qquad \qquad \qquad \phi =\phi'\,  \cos \epsilon  -\omega'\, \sin \epsilon\,,
\nonumber\\
&& \qquad{\rm with} \qquad \Pi_{\omega \phi} =\frac{1}{2}\,\Big(\Pi_\phi - \Pi_\omega \Big)\,\tan (2\,\epsilon) \,.
\end{eqnarray}
In (\ref{tree-level-V}) we use  a convention where the $\omega$ field has no strangeness content. In the transformed field $\omega'$ 
the mixing angle $\epsilon$ is a direct measure for the latter. 

While at order $Q^2$ the mass term contribution proportional to $M^2$ from $\mathcal L_2^{(2)}$ in (\ref{def-L22}) 
does not predict a mixing of the $\omega$ and $\phi$ meson, this is no longer true once the 
counter terms relevant at $Q^4$ are considered. The leading mixing effect is induced by the parameters $e_2, b_2$ and $b_3$. 
The leading terms proportional to the square of a quark mass, $c_1$ and $c_2$, do not induce mixing effects
\footnote{Also the subleading 
parameters $c_{3}, c_4$ and $c_7$ in (\ref{def-extra-terms}) do not do so. With
\begin{eqnarray}
 c_6 +2\,c_8 +4\,c_9 + 8\,c_{10 } = 0\,,\quad  c_6 +c_8 +2\,c_9 + 2\,c_{10 } = 0\,, \quad 
 2\,c_5+3\,c_8 + 8\,c_{10 } = 0\,,
 \label{res-no-mixing}
\end{eqnarray} 
conditions are obtained that exclude any mixing effect.}.

\begin{table}[t]
\begin{tabular}{l||c|c|c||c|c}
\;\;V\;\;& \;\;Q  \;\;          & $G^{(S)}_{VQ}$                          & $G^{(T)}_{VQ}$                         &\;\; R\;\;& $G_{VR}^{(T)}$                                                      \\
\hline 
	& $\pi$                 & $\frac{1}{2}\,g_1 + g_2$                & $g_3 - \frac{2}{3}\,g_5$               & $\rho$   & $- \frac{5}{ 6}\,(g_6-g_7)- \frac{13}{24}\,g_8+ \frac{1}{6}\,g_9$   \\
$\rho\,\rho$  & $K$                   & $\frac{1}{2}\,(g_1 + g_2)$              & $\frac{1}{3}\,(2\,g_3 - g_5)$          & $K^*$    & $- \frac{3}{4}\,g_6 + \frac{5}{12}\,(g_7-g_8)+ \frac{1}{12}\,g_9$   \\ 
	& $\eta$                & $\frac{1}{6}\,g_1$                      & $\frac{1}{9}\,g_3$                     & $\omega$ & $- \frac{2}{3}\,g_6 - \frac{7}{24}\,g_8$                            \\ 
	&                       &                                         &                                        & $\phi$   & $0$                                                                 \\ 
\hline
	& $\pi$                 & $\frac{3}{2}\,g_1$                      & $g_3$                                  & $\rho$   & $- 2\,g_6 - \frac{7}{8}\,g_8$                               \\ 
$\omega\,\omega$& $K$                   & $\frac{1}{2}\,(g_1 + g_2)$              & $\frac{1}{3}\,(2\,g_3 - g_5)$          & $K^*$    & $- \frac{3}{4}\,g_6 + \frac{5}{12}\,(g_7-g_8)+ \frac{1}{12}\,g_9$   \\ 
	& $\eta$                & $\frac{1}{6}\,g_1$                      & $\frac{1}{9}\,g_3$                     & $\omega$ & $- \frac{2}{3}\,g_6 - \frac{7}{24}\,g_8$                       \\ 
	&                       &                                         &                                        & $\phi$   & $0$                                                                 \\
\hline
	& $\pi$                 & $\frac{3}{8}\,(g_1 + g_2)$              & $\frac{1}{4}\,(2\,g_3 -g_5)$           & $\rho$   & $-\frac{9}{16}\,g_6 + \frac{5}{16}\,(g_7-g_8)+ \frac{1}{16}\,g_9$   \\  
$K^* K^*$   & $K$                   & $\frac{3}{4}\,(g_1 + g_2)$              & $g_3-\,\frac{1}{2}\,g_5$               & $K^*$    & $-\frac{9}{ 8}\,g_6 + \frac{5}{ 8}\,(g_7-g_8)+ \frac{1}{ 8}\,g_9$   \\ 
	& $\eta$                & $\frac{1}{24}\,g_1 + \frac{3}{8}\,g_2$  & $\frac{5}{18}\,g_3 - \frac{1}{4}\,g_5$ & $\omega$ & $-\frac{3}{16}\,g_6 + \frac{5}{48}\,(g_7-g_8)+ \frac{1}{48}\,g_9$   \\ 
	&                       &                                         &                                        & $\phi$   & $- \frac{3}{8}\,g_6 + \frac{5}{24}\,(g_7-g_8)+ \frac{1}{24}\,g_9$   \\ 
\hline
	& $\pi$                 & $0$                                     & $0$                                    & $\rho$   & $0$                                            \\
$\phi\, \phi$  & $K$                   & $g_1 + g_2$                             & $\frac{2}{3}\,(2\,g_3 - g_5)$          & $K^*$    & $- \frac{3}{2}\,g_6 + \frac{5}{6}\,(g_7-g_8)+ \frac{1}{6}\,g_9$   \\ 
	& $\eta$                & $\frac{2}{3}\,g_1$                      & $\frac{4}{9}\,g_3$                     & $\omega$ & $0$                                            \\
	&                       &                                         &                                        & $\phi$   & $- \frac{4}{3}\,g_6 - \frac{7}{12}\,g_8$   \\  
\hline
            & $\pi$     & $0$                           & $0$                      &$\rho$& $0$                                             \\

$\omega\,\phi$& $K$       &$\frac{1}{\sqrt{2}}\,(g_1-g_2)$&$\frac{\sqrt{2}}{3}\, g_5$& $K^*$& $-\frac{7}{6\sqrt{2}}g_6 - \frac{5}{6\sqrt{2}}\,(g_7 + g_8)-\frac{1}{6\sqrt{2}}g_9$\\
            & $\eta$    & $0$                           & $0$                      &$\omega$& $0$                                             \\
            &                       &                                           &                                      &$\phi$  & $0$                                             \\
\hline	
\end{tabular}
\caption{The coupling constants $G^{(S)}_{VQ}$ and $G^{(T)}_{VQ}$ as introduced in (\ref{Pi-V-tadpole}). }
\label{tab:V-tadpole}
\end{table}

We continue with a coherent documentation of the one-loop contributions to the vector meson self energies which is decomposed into a tadpole and a bubble contribution with
\begin{eqnarray}
 \Pi_V^{\rm loop} =  \Pi_V^{\rm tadpole}+  \Pi_V^{\rm bubble}\,.
 \label{def-loop}
\end{eqnarray}
It is convenient to 
start with terms that result from two-body vertices involving two vector meson fields in (\ref{def-L22}) and (\ref{def-L24}). 
Tadpole structures arise where either a 
Goldstone boson tadpole $\bar I_Q$ with $Q \in [8]$ or a vector meson tadpole $\bar I_V$ with $V \in[9]$ is formed. We express our result 
\begin{eqnarray}
&& \Pi^{\rm tadpole}_{V\in [9]}  =  \frac{1}{4\,f^2}\sum_{Q\in [8]}\Big\{-2\,m_Q^2\,\Big(4\,G^{(S)}_{VQ}+ G^{(T)}_{VQ}\Big)\,\bar I_Q 
- 4\,G^{(T)}_{VQ}\,\bar I_Q^{(2)}  \Big\}
\nonumber\\
&& \qquad \quad \; +\,\frac{3}{4\,f^2}\,\sum_{R\in [9]}\,128\,G_{VR}^{(T)}\,M^2\,\bar I_R\,,
\label{Pi-V-tadpole}
\end{eqnarray}
with the tadpole function
\begin{eqnarray}
\bar I_Q = \frac{m_Q^2}{(4\,\pi)^2}\,\log \frac{m_Q^2}{\mu^2}\,, \qquad \qquad \bar I^{(2)}_Q = \frac{1}{4}\,m_Q^2\, \bar I_Q \,,
\label{def-IQ2}
\end{eqnarray}
recalled in its infinite volume limit.  The vector  meson tadpole $\bar I_R$ follows from $\bar I_Q$ with the replacement $m_Q \to M_R$.
Our derivation of (\ref{Pi-V-tadpole}) is valid in a finite box, where different species of tadpole integrals 
may occur. We use the notations of  \cite{Lutz:2014oxa} with $\bar I^{(2)}_Q$ and $\bar I_Q$, for which it follows 
the infinite volume limit (\ref{def-IQ2}). For the finite volume case the standard result as for instance shown in \cite{Lutz:2014oxa} 
should be used. Note that analogous terms in the vector meson tadpoles are not resolved in this work. Here for typical QCD lattices   
the finite volume effects are negligible, being suppressed by factors $ e^{-L\,M_V}$ with $L^3$ the volume of the considered box.  
The coefficient matrices $G^{(S)}_{VQ}$, $G^{(T)}_{VQ}$  and $G_{VR}^{(T)}$ are detailed 
in Tab. \ref{tab:V-tadpole}.

There remain the bubble loop diagrams built in terms of the three-point vertices introduced in (\ref{def-L23}). The computation requires a further set of coupling 
constants $G_{QP}^{(V)}$, $G_{R\,T}^{(V)}$ and $G_{QR}^{(V)}$  that specify the strength of a three-point vertex in a given isospin projection. Like in the 
previous section we use $P,Q\in [8]$ for the isospin multiplets of the Goldstone bosons and $V,R,T \in [9]$ for the ones of the vector mesons. The corresponding 
coefficients are collected in Tab. \ref{tab:V-Clebsch}. Our result
\begin{eqnarray}
&& \Pi^{\rm bubble}_{V\in [9]}  =  \sum_{Q,P\in [8]} 
\Bigg(\frac{G_{QP}^{(V)}}{2\,f}\Bigg)^2 \Bigg\{ -\frac{1}{4}\,\Big(m_P^2-m_Q^2\Big)^2\,\Delta I_{QP} 
\nonumber\\
&& \qquad \qquad \qquad   \quad -\,\frac{1}{4}\,M_V^2\,\Big( \bar I_Q + \bar I_P\Big)
-  \frac{1}{4}\,M_V^2\,\Big( M_V^2 -2\, (m_P^2+ m_Q^2)\Big)\,I_{QP} \Bigg\}
\nonumber\\
&& \qquad \quad + \sum_{Q\in [8] ,\,R\in [9]} \,\Bigg(\frac{G_{QR}^{(V)} }{2\,f}\Bigg)^2
\Bigg\{ - \frac{1}{4}\,\alpha^{V}_{QR}\,
\Big( M_R^2-m_Q^2\Big)^2\,\Delta I_{QR} -\frac{1}{4}\, \alpha^{V}_{QR}\,M_V^2\,\bar I_R 
 \nonumber\\
&& \qquad \qquad \qquad   \quad  +\, \beta^{V}_{QR}\,m_Q^2\,\bar I_Q  + \delta^{V}_{QR}\,\bar I_Q^{(2)}  - \frac{1}{4}\,\alpha^{V}_{QR}\,
M_V^2\,\Big( M_V^2- 2\,(m_Q^2+M_R^2)\Big)\,I_{QR} \Bigg\} 
\nonumber\\
&& \qquad \quad  + \sum_{R,T\in [9]} \Bigg(\frac{G_{R \,T}^{(V)}}{2\,f}\Bigg)^2
\Bigg\{ - \frac{1}{4}\,\alpha_{R\,T}^{V}\,\Big( M_R^2-M_T^2\Big)^2 \Delta I_{RT}
+ \beta_{RT}^{V}\,M_V^2\,\bar I_R 
\nonumber\\
&& \qquad \qquad \qquad \quad + \,\delta_{RT}^{V}\, M_V^2\,\bar I_T
-\, \frac{1}{4}\,\alpha_{R\,T}^{V}\,M_V^2\Big( M_V^2-2\,(M_R^2+M_T^2)\Big)\,I_{R\,T}\Bigg\}\,,
\nonumber\\ \nonumber\\
&& \alpha^{V}_{QR} = \frac{ (M_R^2+M_V^2)^2 }{4\, M_R^2\,M_V^2}  \,, \qquad \qquad \qquad  \alpha^{V}_{R\,T} = M^4\,\frac{M_R^2+M_V^2+M_T^2}{3\,M_R^2\,M_V^2\,M_T^2}
\nonumber\\
&&  \beta_{QR}^{V} =- \frac{ (M_R^2+ M_V^2)^2 + 4\,m_Q^2 \,(M_R^2 - M_V^2 )}{16\, M_R^2\, m_Q^2}
 \,,  \qquad  \qquad \beta^{V}_{R\,T} = M^4\,\frac{8\,M_R^2-M_T^2 -M_V^2}{12\,M_R^2\,M_V^2\,M_T^2}\,,
\nonumber\\
&& \delta_{QR}^{V} = \frac{2\, M_R^2+M_V^2}{8\, M_R^2}\,, \qquad \qquad \qquad \;\,\delta^{V}_{R\,T} = M^4\,\frac{8\,M_T^2-M_R^2 -M_V^2}{12\,M_R^2\,M_V^2\,M_T^2}\,,
\label{Pi-vector}
\end{eqnarray}
is expressed in terms the tadpole integrals $\bar I_Q, \bar I_P$ and $\bar I_R, \bar I_T$ and the scalar bubble functions $I_{QP}, \Delta I_{QP}$, $ I_{QR}, \Delta I_{QR}$ and 
$I_{R\,T}, \Delta I_{R\,T}$. We specify the generic case with $QR$  for the infinite volume limit
\begin{eqnarray}
&& I_{QR} = \bar I_{QR} - \frac{\bar I_R}{M_R^2}\,, \qquad \qquad \qquad \qquad \Delta I_{QR} = I_{QR} + \frac{\bar I_Q - \bar I_R}{m_Q^2- M_R^2}\,,
 \nonumber\\
&& \bar I_{QR} = \frac{1}{16\,\pi^2}
\left\{ 1-\frac{1}{2} \left( 1+ \frac{m_Q^2-M_R^2}{M_V^2} \right) \,\log \left( \frac{m_Q^2}{M_R^2}\right)
\right.
\nonumber\\
&& \;\quad \;\,+\left.
\frac{p_{Q R}}{M_V}\,
\left( \log \left(1-\frac{M_V^2-2\,p_{Q R}\,M_V}{m_Q^2+M_R^2} \right)
-\log \left(1-\frac{M_V^2+2\,p_{Q P}\,M_V}{m_Q^2+M_R^2} \right)\right)
\right\} \, ,
\nonumber\\
&& p_{Q R}^2 =
\frac{M_V^2}{4}-\frac{m_Q^2+M_R^2}{2}+\frac{(m_Q^2-M_R^2)^2}{4\,M_V^2}  \,.
\label{def-bubble}
\end{eqnarray} 
Corresponding expressions that hold for the scalar bubble in a finite box can be taken from \cite{Lutz:2014oxa}. The loop functions $I_{QP}$ and 
$I_{R\,T}$ follow from (\ref{def-bubble}) by appropriate replacements of the masses. Explicit expressions appropriate for the finite box case 
can be taken from \cite{Lutz:2014oxa}.

\begin{table}[t]
\begin{tabular}{lll}
$G_{\pi\pi}^{(\rho)} = \frac{1}{\sqrt{3}}\,h_1$                           &  $G_{KK}^{(\rho)} = \frac{1}{\sqrt{6}}\,h_1 \qquad \qquad $    & $G_{KK}^{(\omega)} = \frac{1}{\sqrt{6}}\,h_1$ \\
$G_{\pi K}^{(K^*)} = \frac{1}{\sqrt{8}}\,h_1$                           &  $G_{K\eta}^{(K^*)} = \frac{1}{\sqrt{8}}\,h_1$                      & $G_{KK}^{(\phi)} = - \frac{1}{\sqrt{3}}\,h_1$ \\
\\ \hline
$G_{\pi\omega}^{(\rho)} = 2\,\sqrt{\frac{2}{3}}\,h_2\qquad \qquad  $      & $G_{\pi\rho}^{(\omega)} = 2\,\sqrt{2}\,h_2$                          &  $G_{\pi K^{*}}^{(K^{*})} = \sqrt{2}\,h_2 =G_{K\rho}^{(K^{*})} $     \\
$G_{\eta\rho}^{(\rho)} = \frac{2}{3}\,\sqrt{2}\,h_2$                      & $G_{\eta\omega}^{(\omega)} = \frac{2}{3}\,\sqrt{2}\,h_2$             &  $G_{K\omega }^{(K^{*})} = \sqrt{\frac{2}{3}}\,h_2$           \\
$G_{KK^{*}}^{(\rho)} = 2\,\sqrt{\frac{2}{3}}\,h_2$                  & $G_{KK^{*}}^{(\omega)} = 2\,\sqrt{\frac{2}{3}}\,h_2$           &  $G_{\eta K^{*}}^{(K^{*})} = -\frac{\sqrt{2}}{3}\,h_2$    \\
$G_{K K^{*}}^{(\phi)} = \frac{4}{\sqrt{3}}\,h_2$                    & $G_{\eta\phi}^{(\phi)} = -\frac{4}{3}\,\sqrt{2}\,h_2$                &  $G_{K\phi}^{(K^{*})} = \frac{2}{\sqrt{3}}\,h_2$   \\
\\ \hline
$G_{\rho\rho}^{(\rho)} = 6\,\sqrt{2}\,h_3$                                & $G_{K^*K^*}^{(\rho)} = 6\,h_3$                                 & $G_{K^*K^*}^{(\phi)} = -6\,\sqrt{2}\,h_3$ \\
$G_{\rho K^*}^{(K^*)} = 3\,\sqrt{3}\,h_3$                                & $G_{\omega K^*}^{(K^*)} = 3\,h_3$                                    & $G_{\phi K^*}^{(K^*)} = -3\,\sqrt{2}\,h_3= -G_{K^*K^*}^{(\omega)}/\sqrt{2}$ \\
\end{tabular}
  \caption{The coupling constants for vector mesons 
  $G_{QR}^{(V)}$ with $V,\,R\in [9]$ and $P,Q \in [8]$ defined with respect to 
  isospin states. Coupling constants that vanish due to G-parity 
  considerations are not shown. It holds $G^{(V)}_{P\,Q}=G^{(V)}_{QP}$ and $G^{(V)}_{R\,T}=G^{(V)}_{TR}$. }
 \label{tab:V-Clebsch}
 \end{table}

The one-loop self energy (\ref{Pi-vector})  will be analyzed in the following. In particular its renormalization is scrutinized. We begin with 
the chiral limit of the vector meson masses. The reader may ask about the relevance of the chiral limit value, $M_\chi$, of the vector meson masses. After all 
in this limit the phase space of the decay of a vector meson into pairs of Goldstone bosons is wide open and one may expect a significant decay width $\Gamma_\chi$. 
However, this is not the case.  One readily obtains the expression 
\begin{eqnarray}
 \Gamma_\chi = \frac{h_1^2\,M^3_\chi}{ 2\,\pi\,(16\,f)^2} \simeq  0.24\,{\rm GeV}\,,
\end{eqnarray}
where the numerical estimate is obtained with $M_\chi = 0.8$ GeV and $f = 92$ MeV. Such a value for the decay width seems compatible with our formal counting scheme $\Gamma_\chi \sim Q^3$ as 
compared to the scaling of the mass $M^2\sim Q^2 $. 

We turn to the vector meson mass in the chiral limit for which we obtain
\begin{eqnarray}
&& M^2_\chi = M^2 + e_1\,M^4 -  \frac{3 \,M^4 }{2 \pi ^2\, f^2} \,\Big(9\, g_6-5 \,g_7+5 \,g_8-g_9\Big) \,\log \frac{M^2}{\mu ^2} 
\nonumber\\
&& \; -\, \frac{h_1^2 \,M^4}{512 \pi ^2 f^2}\left(1-\log \frac{M^2}{\mu^2}\right)
 -\frac{7\, h_2^2 \,M^4 }{144 \pi ^2 f^2} \,\log \frac{M^2}{\mu ^2}+ \frac{27\,h_3^2\,M^4}{64 \pi^2 f^2} \left( 3 -\sqrt{3} \pi +\log \frac{ M^2}{\mu^2} \right)\,,
\label{ren-M}
\end{eqnarray}
where we use $ M_\chi$ for the renormalized mass of the vector mesons in the chiral limit. 
Here we replaced all vector meson masses $M_V,M_R, M_T \to M$ in (\ref{Pi-V-tadpole}, \ref{Pi-vector})
by the leading order expression. In addition all masses of the Goldstone bosons $m_Q, m_P \to 0$ are put to zero.  The unknown parameter $e_1$ 
is needed to render the vector meson masses independent of the renormalization scale. All $\log \mu$ 
terms in (\ref{ren-M}) can be properly balanced by $e_1$.

The result (\ref{ren-M}) is interesting since it gives a first hint 
on the importance of loop corrections to the vector meson masses. As expected the loop corrections are suppressed in the large-$N_c$ limit of QCD.
This is manifest in (\ref{ren-M}) with the scaling behavior $f^2\sim N_c$ and $M,g_n, h_n \sim N_c^0$ and $e_1\sim 1/N_c$. 
Is this formal scaling property supported by corresponding numerical values at $N_c =3$?  At the renormalization scale $\mu =  M= 0.80$ GeV and 
the particular parameter choices  (\ref{parameter-set-I}) we derive the estimate
\begin{eqnarray}
M^2_\chi - M^2  \simeq  e_1\,M^4  -0.118\, M^2 \qquad {\rm at } \qquad \mu=  M \,,
\label{estimate-Mchi}
\end{eqnarray}
which does not depend on the so far unknown low-energy parameters $g_n$. The correction term (\ref{estimate-Mchi}) is comfortable small giving support to the 
assumed dimensional counting rules.

We consider a further physical quantity. The chiral limit of the $\omega- \phi$ mixing parameter is renormalized with  
\begin{eqnarray}
&& \frac{1}{\sqrt{2}} \,\Pi_{\omega \phi} \Big|_{m=0=m_s} =  e_2\,M^4 - \frac{M^4 }{2 \pi^2\, f^2} \,\Big(7 \,g_6+5\, g_7+2 \,g_8+g_9\Big)\,\log \frac{M^2}{\mu ^2}
\nonumber\\
&& \;+\, \frac{h_1^2\,M^4  }{1536 \pi ^2 f^2}\left(1-\log \frac{M^2}{\mu^2}\right)
 -\frac{h_2^2\, M^4}{48 \pi ^2 f^2}\ln \left(\frac{M^2}{\mu ^2}\right)-\frac{9\,h_3^2\,M^4}{64 \pi^2 f^2} \left( 3 -\sqrt{3} \pi +\log \frac{ M^2}{\mu^2} \right) \,,
 \end{eqnarray}
where we replaced again all vector meson masses $M_V,M_R, M_T \to M$ in (\ref{Pi-V-tadpole}, \ref{Pi-vector})
by the leading order expression \footnote{Upon the 
replacement $$G^{(V)}_{\cdots}\,G^{(V)}_{\cdots} \to G^{(\omega)}_{\cdots}\,G^{(\phi)}_{\cdots}\,,$$
the bubble loop contribution to $\Pi_{\omega \phi}$ is readily constructed from (\ref{Pi-vector}). }. In addition all masses of the Goldstone bosons $m_Q, m_P \to 0$ are put to zero. 
The parameter $e_2$ makes the mixing angle scale invariant. 
The loop correction is suppressed by $1/N_c$ as expected from the OZI rule. 
Even numerically we obtain with
\begin{eqnarray}
\frac{1}{\sqrt{2}} \,\Pi_{\omega \phi} \Big|_{m=0=m_s} \simeq e_2\,M^4  + 0.039\,M^2 \qquad {\rm at } \qquad \mu= M \,,
\end{eqnarray}
a small contribution to the mixing angle.  

We continue with a study of the scale dependence of the remaining low-energy parameters. First we identify all scale-dependent terms in (\ref{Pi-V-tadpole}, \ref{Pi-vector}) that are proportional to $M^2\,m_{P,Q}^2$. 
Again  $M_V,M_R, M_T \to M$ is used. Such terms define the running of the low-energy parameters $b_i$ as follows
\begin{eqnarray}
&& \mu^2 \frac{\text{d}}{\text{d}\mu^2}\, b_1 =- \frac{9 \, h_1^2 + 8\, h_2^2}{1536 \pi^2 \,f^2}  \,, \qquad  \qquad 
\mu^2 \frac{\text{d}}{\text{d}\mu^2} \, b_2 =  -\frac{9 \, h_1^2 + 104\, h_2^2 }{4608 \pi^2  \,f^2} \,,\qquad 
\nonumber\\
&& \mu^2 \frac{\text{d}}{\text{d}\mu^2} \,b_3 = \frac{  h_1^2 - 8 h_2^2 }{256 \pi^2 \,f^2}  \,.
\label{res-b-run}
\end{eqnarray}
For  the symmetry breaking parameters $c_i$ we observe an interesting phenomenon. 
The scale invariance of the corresponding loop structures can be achieved only, if specific correlations on the symmetry conserving low-energy parameters $g_i$  are imposed. This is seen as follows. 
A priori all scale dependent terms can be balanced only if we activate the N$^2$LO and N$^3$LO counter terms in (\ref{def-extra-terms}). We find
\begin{eqnarray}
&&\mu^2 \frac{\text{d}}{\text{d}\mu^2} \, c_1 = \frac{4 \,g_1 -4 \,g_2 + g_5}{192 \pi^2 f^2} - \frac{ h_2^2 }{768\, \pi^2 f^2}\,, 
\nonumber\\
&&\mu^2 \frac{\text{d}}{\text{d}\mu^2} \, c_2 = \frac{44 \,g_1 -20 \,g_2 +6 \,g_3 +5 \,g_5}{768 \pi^2 f^2} - \frac{11\, h_2^2 }{3072\, \pi^2 f^2}\,, 
\nonumber\\ \nonumber\\
&&M^2\,\mu^2 \frac{\text{d}}{\text{d}\mu^2} \, c_3 = -\frac{68 \,g_1 +36 \,g_2 +26 \,g_3 -9 \,g_5}{2304 \pi^2 f^2} + \frac{17 h_2^2}{9216\, \pi^2 f^2}\,, 
\nonumber\\
&&M^2\,\mu^2 \frac{\text{d}}{\text{d}\mu^2} \, c_4 = - \frac{68 \,g_1 +36 \,g_2 +26 \,g_3 -9 \,g_5}{1152\, \pi^2 f^2} + \frac{17 h_2^2}{4608\, \pi^2 f^2}\,, 
\nonumber\\
&&M^2\,\mu^2 \frac{\text{d}}{\text{d}\mu^2} \,\big( c_5 + c_6 \big) = -\frac{4 \,g_1 -4 \,g_2 + g_5 }{64 \pi^2 f^2} + \frac{ h_2^2}{256\, \pi^2 f^2}\,,  
\nonumber\\ \nonumber\\
&&M^4\,\mu^2 \frac{\text{d}}{\text{d}\mu^2} \, c_7 = \frac{4 \,g_1 + g_3 }{864\pi^2 f^2} -\frac{ h_2^2}{3456\, \pi^2 f^2}\,, 
\nonumber\\
&&M^4\,\mu^2 \frac{\text{d}}{\text{d}\mu^2} \, c_8 = 0\,, 
\nonumber\\
&&M^2\,\mu^2 \frac{\text{d}}{\text{d}\mu^2} \, c_9 =  0\,.
\label{running-ci}
\end{eqnarray}
From (\ref{running-ci}) we conclude that the hadrogenesis Lagrangian is renormalizeable only if  the following 
two sum rules
\begin{eqnarray}
&&4\,g_1+ g_3 = \frac{1}{4}\,h_2^2\,, \qquad \qquad \qquad  g_5 = g_3 +4\,g_2\,,
\label{sum-rules-gi}
\end{eqnarray}
hold at leading order in the power counting scheme. In fact, we observe that once we insist on those two relations the leading order parameters $c_1$ and $c_2$ remain scale invariant.

\begin{figure}
\centering
\includegraphics[width=0.4\textwidth]{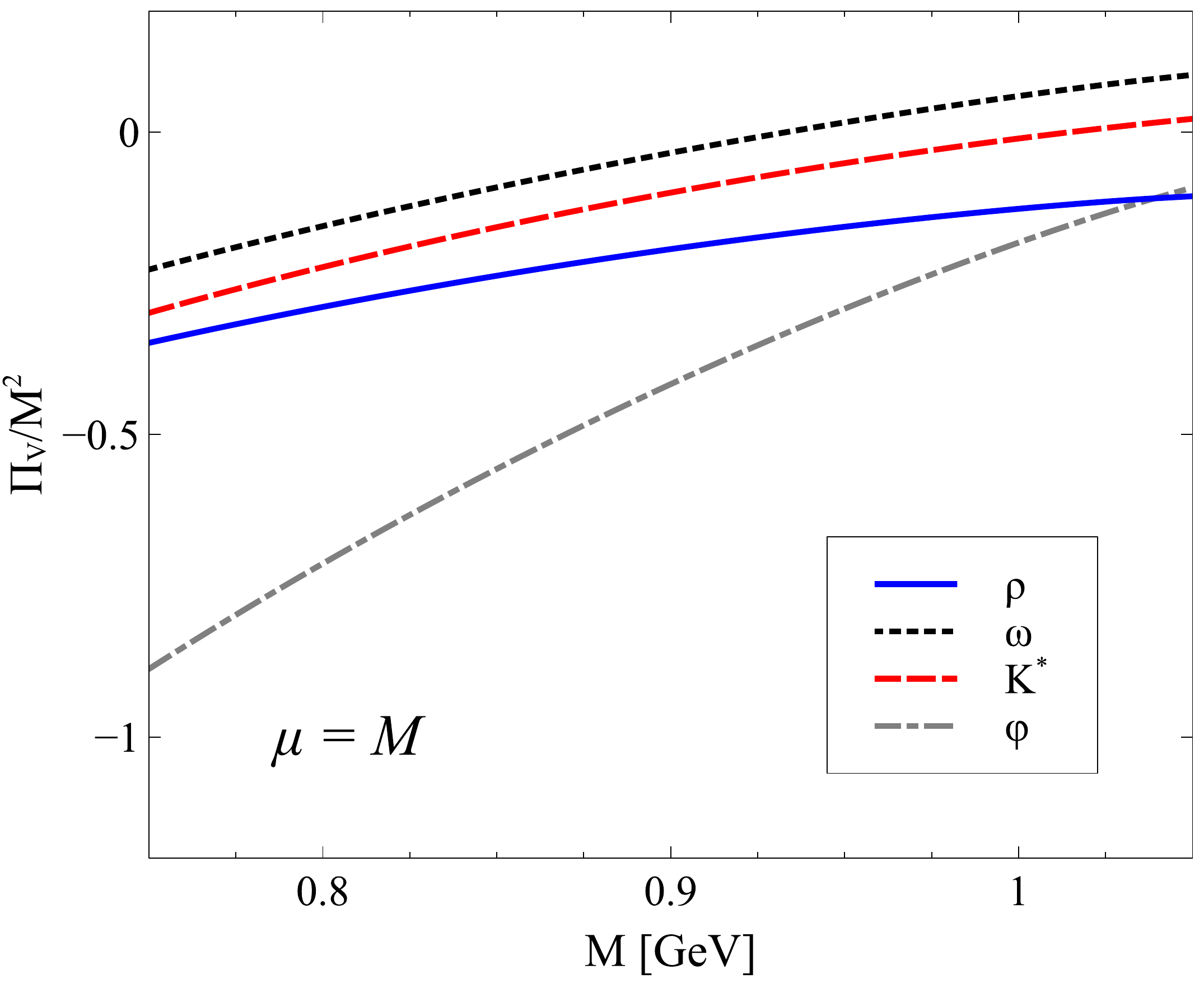}
\caption{The vector meson polarization  $\Pi^{\rm loop}_V/M^2$ are presented as a function of $M$, at $\mu=M$. The plots rely on leading order bare masses for all mesons. }
\end{figure}

In Fig. 1  we further scrutinize the numerical implications of our approach. We computed the polarization tensors for the four vector mesons as a function of the mass
parameter $M $ for the particular renormalization scale $\mu = M$. The  ratios $\Pi^{\rm loop}_V /M^2$ are plotted in order to provide a direct measure for the importance of the loop effects. The physical vector meson masses 
are reproduced upon a suitable choice of the low-energy parameters determining the size of the tree-level contributions at order $Q^4$. 
Therefore the size of the plotted ratios is  a direct measure for the naturalness of such low-energy parameters. Our results rely significantly on the consistency 
relations (\ref{sum-rules-gi}). As a consequence there does not remain any residual dependence on any of the unknown parameters $g_n$. This is a particular property of the scenario 
with $\mu =M$.

Within the range of expected values  $M_\rho < M < M_\phi$ all shown ratios are systematically smaller than one, however quite large for the $\phi$ meson at the lower bound of $M$. This together 
with the inverted pattern of the loop sizes for the different vector mesons would cause large low-energy parameters, possibly in conflict with the naturalness assumption. 
We note that the loop contributions are dominated largely by $h_2$ (see also \cite{Bijnens:1996nq,Bijnens:1997ni,Leinweber:2001ac}). 
The source of this effect is readily traced. It is a consequence of improperly approximated phases space factors $p^2_{QR}$ using the replacement $M_{V,R,T } \to M$. This is a phenomenon known already 
from $\chi$PT studies of baryon masses \cite{Semke2005,Lutz:2014oxa,Lutz:2018a}. An efficient remedy is the use of physical masses in the loop function \cite{Semke2005,Lutz:2014oxa,Lutz:2018a,Guo:2018a}. 
The immediate concern is acknowledged: can any such scheme be scale invariant? In recent works \cite{Lutz:2018a,Guo:2018a} a method was suggested that indeed leads to scale invariant results.

\begin{figure}
\centering
\includegraphics[width=0.8\textwidth]{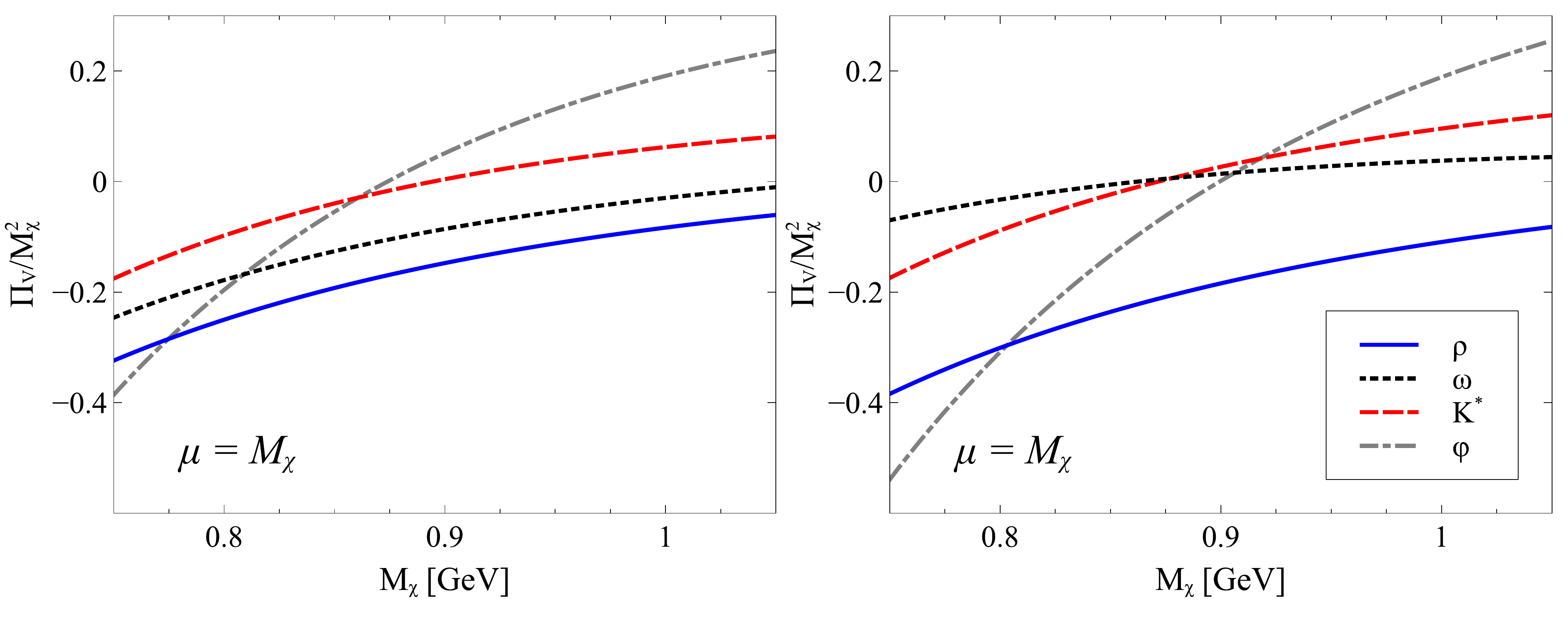}
\caption{The vector meson polarization  $\Pi^{\rm loop}_V/M_\chi^2$ are presented as a function of $M_\chi$, at $\mu=M_\chi$ where physical values for all meson masses are assumed.
While the left-hand plot show results at vanishing $\omega-\phi$ mixing angles the right-hand plot illustrates 
the effect of non-vanishing $\omega-\phi$ mixing angles. Distinct values for the mixing angles at the $\omega$ and $\phi$ meson poles 
are assumed as explained in the text below. }
\end{figure}

In the following we will adapt the formalism \cite{Lutz:2018a,Guo:2018a} to our case at hand. In a first step we renormalize away all vector meson tadpole contributions with
\begin{eqnarray}
&& \bar I_R \to 0 \,,\qquad \qquad \qquad \qquad \quad \;\; \bar I_T \to 0 \,, 
\nonumber\\
&& I_{QR} = \bar I_{QR} - \frac{\bar I_R}{M_R^2}\to \bar I_{QR} \,,\qquad I_{R\,T} = \bar I_{R\,T} - \frac{1}{2}\,\Bigg( \frac{\bar I_R}{M_R^2} + \frac{\bar I_T}{M_T^2} \Bigg) \to  \bar I_{R\,T}\,,
\nonumber\\
&& I_{QP} = \bar I_{QP} - \frac{\bar I_V}{M_V^2}\to \bar I_{QP} \,,
\label{def-sub1}
\end{eqnarray}
where the residual objects $\bar I_{QR}$, $\bar I_{R\,T}$ and $\bar I_{QP}$ are scale invariant by construction. We emphasize that a subtraction scheme for the loop functions 
if performed at the level of the Passarino Veltman functions  is symmetry conserving \cite{Lutz2000,Semke2005,Lutz:2014oxa}. As long as 
there is an unambiguous prescription how to represent all one-loop functions in terms of the Passarino Veltman functions we do not expect any 
violation of chiral Ward identities. In a second step we need to set up a power counting for physical masses. The crucial relations  
\begin{eqnarray}
 m_Q^2 \sim Q^2\,,\qquad \quad\frac{m_Q^2-m_P^2}{m_Q^2}\sim Q^0\,,\qquad \quad M_V^2\sim Q^2\,,\qquad \quad  \frac{M_R^2- M_V^2}{M_R^2} \sim Q^2 \,,
 \label{def-count}
\end{eqnarray}
are to be applied to all scale dependent terms. With (\ref{def-sub1}) and (\ref{def-count}) we obtain the result
\begin{eqnarray}
&& \Pi^{\rm bubble}_{V\in [9]}  =  \sum_{Q,P\in [8]} 
\Bigg(\frac{G_{QP}^{(V)}}{2\,f}\Bigg)^2 \Bigg\{ -\frac{1}{4}\,\Big(m_P^2-m_Q^2\Big)^2\,\Delta I_{QP} 
\nonumber\\
&& \qquad \qquad \qquad   \quad -\,\frac{1}{4}\,M_V^2\,\Big( \bar I_Q + \bar I_P\Big)
-  \frac{1}{4}\,M_V^2\,\Big( M_V^2 -2\, (m_P^2+ m_Q^2)\Big)\,\bar I_{QP} \Bigg\}
\nonumber\\
&& \qquad \quad + \sum_{Q\in [8] ,\,R\in [9]} \,\Bigg(\frac{G_{QR}^{(V)} }{2\,f}\Bigg)^2
\Bigg\{ - \frac{1}{4}\,\alpha^{V}_{QR}\,
\Big( M_R^2-m_Q^2\Big)^2\,\Delta I_{QR} 
 \nonumber\\
&& \qquad \qquad \qquad   \quad  -\, \frac{1}{4}\,M^2_V  \, \bar I_Q  
+ \frac{3}{8}\,\bar I_Q^{(2)}  
- \frac{1}{4}\,\alpha^{V}_{QR}\,
M_V^2\,\Big( M_V^2- 2\,(m_Q^2+M_R^2)\Big)\,\bar I_{QR} \Bigg\} 
\nonumber\\
&& \qquad \quad  + \sum_{R,T\in [9]} \Bigg(\frac{G_{R \,T}^{(V)}}{2\,f}\Bigg)^2
\Bigg\{ - \frac{1}{4}\,\alpha_{R\,T}^{V}\,\Big( M_R^2-M_T^2\Big)^2 \Delta I_{RT}
\nonumber\\
&& \qquad \qquad \qquad \quad -\, \frac{1}{4}\,\alpha_{R\,T}^{V}\,M_V^2\Big( M_V^2-2\,(M_R^2+M_T^2)\Big)\,\bar I_{R\,T}\Bigg\} + {\mathcal O} \left(Q^6 \right)\,.
\label{Pi-vector:B}
\end{eqnarray}
We note that there are three terms left only that come with a scale dependence in (\ref{Pi-vector:B}). A simplification arises in the infinite volume limit with
$m_Q ^2 \bar I_Q \to 4\,\bar I^{(2)}_Q $. The term proportional to $\bar I^{(2)}_Q$ in (\ref{Pi-vector:B}) is canceled identically by 
corresponding terms in (\ref{Pi-V-tadpole}) if for the coupling constants $g_{1-5}$ our sum rules (\ref{sum-rules-gi}) are imposed. We consider the terms 
proportional to $\bar I_Q$ or $\bar I_P$ in (\ref{Pi-vector:B}). Their 
scale dependence can be balanced with
\begin{eqnarray}
&& \mu^2 \frac{\text{d}}{\text{d}\mu^2}\, b^r_1 =-\frac{9 \, h_1^2 + 8\, h_2^2}{4608 \pi^2 \, f^2}  \,, \qquad  \qquad 
\mu^2 \frac{\text{d}}{\text{d}\mu^2} \, b^r_2 = -\frac{9 \, h_1^2 + 104\, h_2^2 }{13824 \pi^2\, f^2}\,,\qquad 
\nonumber\\
&& \mu^2 \frac{\text{d}}{\text{d}\mu^2} \,b^r_3 = \frac{  h_1^2 - 8 h_2^2 }{768 \pi^2\, f^2}   \,,
\label{res-br-run}
\end{eqnarray}
where we point at the factor changes in (\ref{res-br-run}) as compared to (\ref{res-b-run}). With $b_i^r$ we denote the low-energy parameters that result in the scheme where the 
Passarino Veltman subtractions (\ref{def-sub1}) are imposed. It is noted that, at first, such counter terms cancel the scale dependence only if the meson masses $m_P$ and $m_Q$ in 
(\ref{Pi-vector:B}) are replaced by their leading order representation as given by the Gell-Mann-Oakes-Renner relations. In addition the replacement $M_V \to M_\chi$ is needed.  However, 
following the previous work \cite{Lutz:2018a,Guo:2018a} we may recast the relevant quark mass terms in (\ref{tree-level-V}) into structures proportional to $m_\pi^2$, $m_K^2$ and $m^2_\eta$, where now the meson masses are not 
constrained by the Okubo relation $m_\eta^2= (4\, m_K^2-m_\pi^2)/3$ any longer. The particular form as detailed in Tab. \ref{tab:zz1} is dictated by the request of scale independence. 
Such a rewrite is unambiguous. It is readily constructed in terms of the convenient linear  combinations
\begin{eqnarray}
 \tilde b_1 = \frac{6\,b^r_1 - b^r_3}{160}\,, \qquad \qquad 
 \tilde b_2 =  \frac{3\,b^r_2}{8}\,, \qquad \qquad 
 \tilde b_3 =  \frac{96\,b^r_1 + 144\,b^r_3}{7680}\,,
\end{eqnarray}
with the scale dependence of $ \tilde b_1$ and $\tilde b_3$ being determined by either $h_2$ or $h_1$. 
We assure that with the rewrite of Tab. \ref{tab:zz1} our vector meson polarization tensors as implied by (\ref{Pi-vector:B}) are scale invariant strictly.

\begin{table}[t]
\setlength{\tabcolsep}{1.5mm}
\renewcommand{\arraystretch}{1.}
\begin{center}
\begin{tabular}{c|cc}\hline
$V$                                 & $V=\rho$\,                                                   & $V= \omega$                                \\  \hline \hline
$M_V^2\,m_\pi^2$                    & $23 \,\tilde b_1 + 3 \,\tilde b_2 + 9 \,\tilde b_3 $         & $3\, (-3 \,\tilde b_1 + \,\tilde b_2 + 35 \,\tilde b_3)$        \\
$M_V^2\,m_K^2$                      & $4 \,(\tilde b_1 + \,\tilde b_2 - \,\tilde b_3) $            & $4 \,(\tilde b_1 + \,\tilde b_2 - \,\tilde b_3)$        \\
$M_V^2\,m_\eta^2$                   & $ -3 \,\tilde b_1 + \,\tilde b_2 + 3 \,\tilde b_3$         & $-3 \,\tilde b_1 + \,\tilde b_2 + 3 \,\tilde b_3 $          \\ \hline

$V$                                 & $V=K^*$\,                                                    & $V= \phi$                                \\  \hline \hline
$M_V^2\,m_\pi^2$                    & $3 \,(\tilde b_1 + \,\tilde b_2 - \,\tilde b_3)$             & $3 \,(-3 \,\tilde b_1 + \,\tilde b_2 - 13 \,\tilde b_3)$        \\
$M_V^2\,m_K^2$                      & $4\, (3 \,\tilde b_1 + \,\tilde b_2 + 5 \,\tilde b_3)$         & $4 \,(5 \,\tilde b_1 + \,\tilde b_2 + 11 \,\tilde b_3)$        \\
$M_V^2\,m_\eta^2$                   & $ 9 \,\tilde b_1 + \,\tilde b_2 - 9 \,\tilde b_3$            & $-3 \,\tilde b_1 + \,\tilde b_2 + 51 \,\tilde b_3 $          \\ \hline
\end{tabular}
\caption{A rewrite of some terms in (\ref{tree-level-V}). With $M_V \to M$ and using the Gell-Mann-Oakes-Renner relations for the meson masses the original expressions are recovered 
identically. We}
\label{tab:zz1} 
\end{center}
\end{table}
We wish to introduce yet a further additional subtraction for later convenience. With 
\begin{eqnarray}
\bar I_{QP} \to   \bar I_{QP} -\frac{1}{(4\pi)^2} \,, \qquad \qquad \bar I_{R\,T} \to   \bar I_{R\,T} -\frac{1- \pi/\sqrt{3}}{(4\pi)^2} \,, 
\label{def-sub2}
\end{eqnarray}
we avoid any renormalization of the chiral limit mass value of the vector mesons. 
All low-energy parameters that result in the scheme where the 
Passarino Veltman subtractions (\ref{def-sub1}) and (\ref{def-sub2}) are applied receive an upper index $'{\rm r\,}'$ as to discriminate them from the 
bare parameters used initially. For instance we have $ M_\chi = M^{\rm r} $ and $e_1^{\rm r} = 0$.

In the left-hand panel of Fig. 2 we show the ratios $\Pi^{\rm loop}_V/M_\chi^2$ from (\ref{Pi-vector:B}). Here 
physical values for all meson masses together with the subtraction scheme (\ref{def-sub1}, \ref{def-sub2}) are scrutinized. 
The renormalization scale $\mu =M_\chi$ is identified with $M= M_\chi$. The ratios in Fig. 2  for the 
four vector mesons show an improved pattern as compared to the corresponding ratios of Fig. 1. The largest loop correction is obtained for the $\phi$ meson. 
All ratios are reasonably small implying natural sized counter terms.

Before providing a first numerical scenario for the set of low-energy constants 
it should be mentioned that 
in the proposed scheme, where physical masses are used throughout all loop function, the $\omega -\phi$ mixing angle turns energy dependent necessarily. 
We deal with this situation by using two distinct 
mixing angles $\epsilon_\omega$ and $\epsilon_\phi$ which are introduced at the $\omega $ and $\phi$ masses respectively. The mixing angles are then determined by the request that the 
transition polarization tensor
\begin{eqnarray}
\Pi_{\omega' \phi'}(s= m_\omega^2 ) = 0 = \Pi_{\omega' \phi'}(s=m_\phi^2 ) \,,
\end{eqnarray}
as computed in the prime basis vanishes at the $\omega$ meson {\it and } the $\phi$ meson masses. These conditions determine the two mixing angle $\epsilon_\omega$ and $\epsilon_\phi$ in a self consistent 
manner. The arsing pattern for the mixing angles is anticipated with Fig. 3. 
In order to keep the scale invariance of our approach we need to recast the tree-level contribution into the following form
\begin{eqnarray}
&& \Pi^{\rm tree}_{\omega \phi}(s) = \sqrt{2}\,e^{\rm r}_2\,s^2 + \sqrt{2}\,b^{\rm r}_3\,s\, m_K^2 \,  .
\label{tree-level-V-mix}
\end{eqnarray}
We emphasize that the tree-level and loop contributions to the mixing function $\Pi_{\omega' \phi'}(s)$ are evaluated with respect to the $\omega'$ and $\phi'$ fields. This is readily achieved  
in terms of the Clebsch coefficients in Tab. I and Tab. II properly rotated into the $\omega', \phi'$ basis. The case where mixing effects involve a vector meson propagating inside a loop contribution 
requires particular care. A scale invariant treatment arises if only terms proportional to any of the scalar bubble terms are rotated. This is readily justified since 
any residual tadpole term $\bar I_Q$  is not associated with an on-shell vector meson for which one can identify its corresponding mixing angle unambiguously. 

\begin{figure}
\centering
\includegraphics[width=0.5\textwidth]{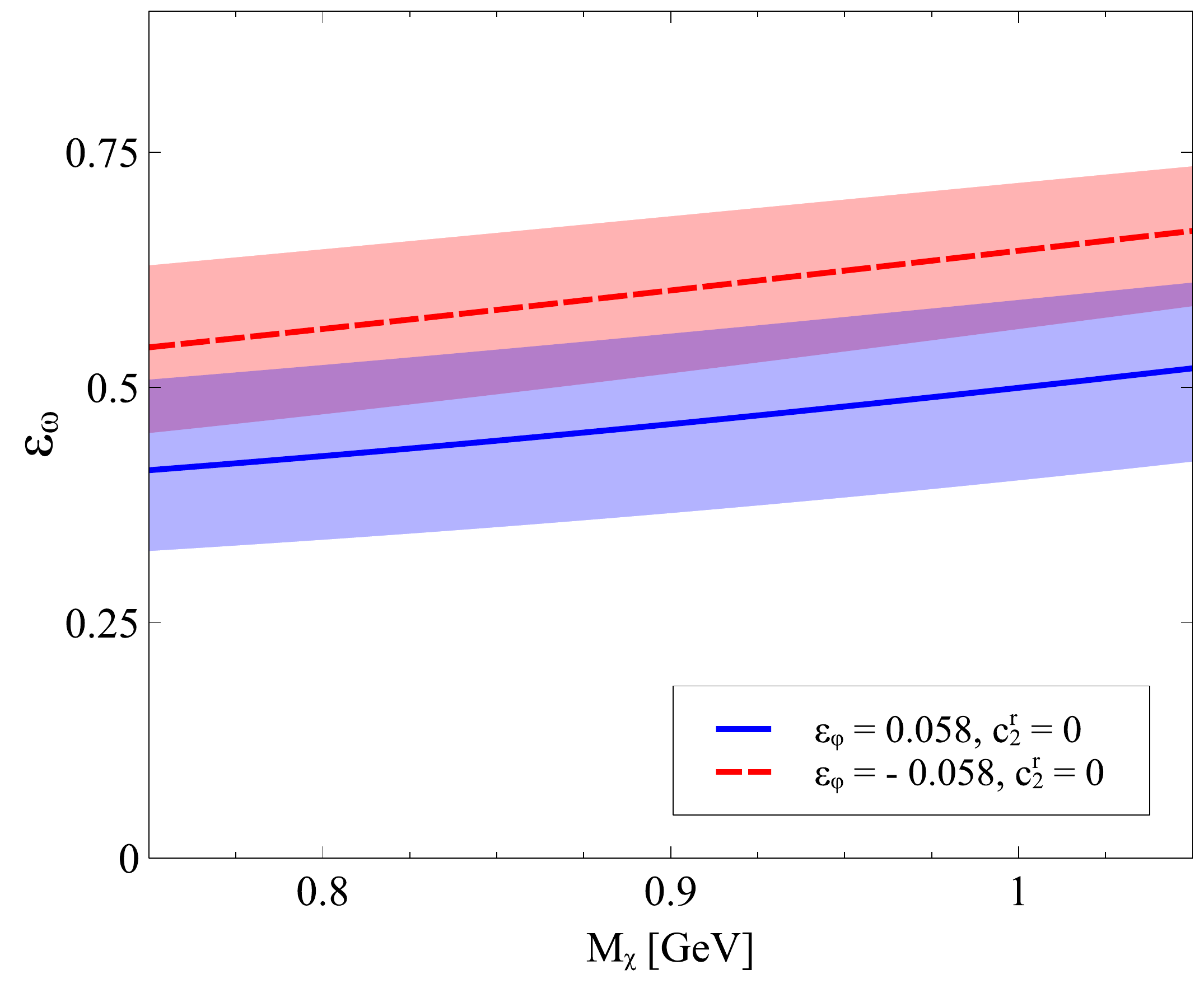}
\caption{The $\epsilon_\omega$ mixing angle derived from the two scenarios $\epsilon_\phi = \pm \,0.058$. The low-energy parameters are adjusted 
as to reproduce the physical meson masses. The bands indicate how $\epsilon_\omega$ changes when $c^{\rm r}_2$ varies from $-1$ to $1$. }
\end{figure}

Based on the scenario using physical meson masses we adjust the low-energy parameters as to recover the physical vector meson masses. At given value for $M$ 
we tune the parameters $ b^{\rm r}_1, b^{\rm r}_2,b^{\rm r}_3$ together with $c^{\rm r}_1$. While the light quark mass $m$ is estimated from the empirical pion mass according to GOR relation, the strange quark mass $m_s$ 
is determined by the empirical ratio $m_s = 27.3 \,m$ \cite{Patrignani:2016xqp}. The parameter $e^{\rm r}_2$ is set such that the  value for the empirical mixing angle 
\begin{equation}
	\epsilon _{\phi} = \pm \,0.058\,.
\label{empirical-eps_phi}
\end{equation}
as  determined from the decay $\phi \to \pi_0\, \gamma$ in \cite{Klingl:1996by} arises. The parameters $c^{\rm r}_2 = 0$ is put to zero initially. 
Its determination requires further empirical input, like it may be provided from QCD lattice simulations of the vector meson masses at non-physical quark masses. 
Note that the effect of $M_\chi$ and  $c^{\rm r}_2$ on the vector meson masses can be discriminated only if data at various choices of the quark masses are considered.
We point out that our estimate for three-point coupling strength $h_2$ needs to be renormalized with 
\begin{eqnarray}
h_2 \simeq (2.33 \pm 0.03) /\cos \epsilon_\omega \,,
\end{eqnarray}
since its previous estimate rests on the decay process $\omega \to \rho \, \pi$ analyzed in the absence of mixing effects \cite{Lutz:2008km,Terschlusen:2012xw}.

\begin{figure}[t]
\centering
\includegraphics[width=0.8\textwidth]{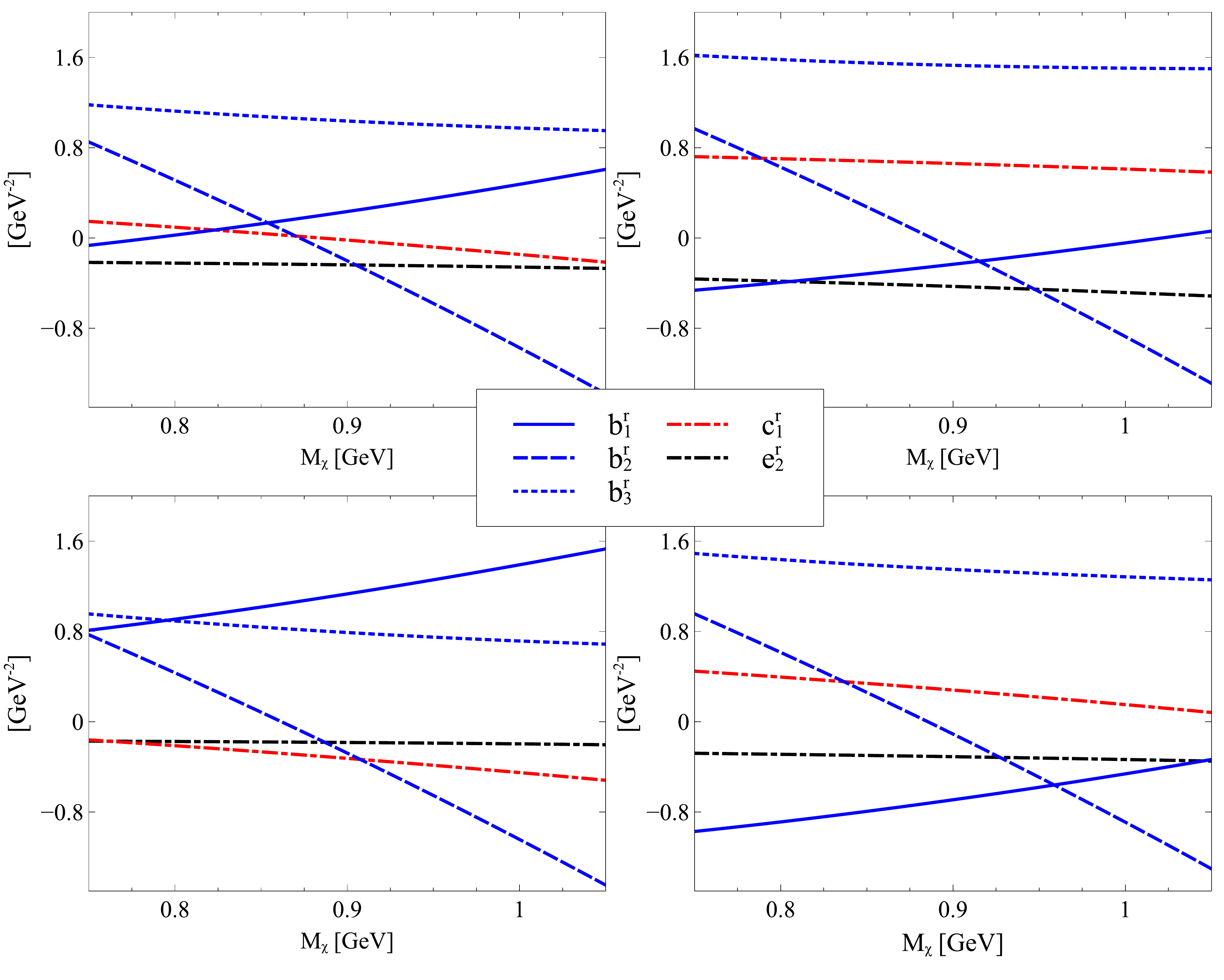}
\caption{The result for the low-energy parameters for a natural range of $M_\chi$ with $\mu = M_\chi$. Physical masses are used. 
While the upper plots  are with $c^{\rm r}_2 = 0$ where  $\epsilon_\phi = 0.058$ (left) and $\epsilon_\phi = -0.058$(right), the lower ones 
follow with $\epsilon_\phi = 0.058$ where $c^{\rm r}_2 = -1$(left) and $c^{\rm r}_2 = 1$(right). }
\end{figure}

In Fig. 3 we present our result for the mixing angle $\epsilon_\omega$ as it results from a fit to the physical masses as described above. 
We observe a significant energy-dependence of the loop-contribution to  $\Pi^{\rm loop}_{\omega\phi}$, in line with the conclusions from previous works \cite{Bijnens:1997ni,Bruns:2013tja}. 
However, we would argue that a proper treatment of the $\omega-\phi$ mixing phenomenon requires a two mixing angle scenario: while the mixing angle may be as small 
as $\epsilon = \epsilon _{\phi} = \pm \,0.058$ at the $\phi$ meson mass, at the $\omega$ meson mass the mixing angle  $\epsilon = \epsilon_\omega$ is an order of magnitude larger. 
We checked with the band widths in Fig. 3 that variations of the form $-1 < c_2 < 1$  do not change this spectacular pattern. It may not come as a surprise that such a large mixing phenomenon 
does mend the form of the loop contributions to the vector meson masses. Indeed as shown in the right-hand plot of Fig. 2 the size of the polarization tensor for the $\omega$ meson is 
affected significantly as compared to the left-hand plot of the same figure that uses $\epsilon_\omega = \epsilon_\phi = 0$. 

In Fig. 4  we show the result for the low-energy parameters in a given range of $M_\chi$, using  physical masses. Note that these were already used in Fig. 3.
For all parameters we obtain naturally sized values.

\clearpage

\section{Goldstone bosons at the one-loop level}
\label{sec:Goldstone-bosons}

We first collect all tree-level contributions to the pseudo-scalar meson masses as implied by $\mathcal L^{(2)}_{2}$ and $\mathcal L^{(2)}_{4}$. 
The well know expression first derived by Gasser and Leutwyler \cite{Gasser:1983yg,Gasser:1984gg} are obtained
\begin{eqnarray}
&& m_\pi^2 =2\,B_0\,m+  \frac{32\,B^2_0\,m}{f^2}\,\Big[ (2\, m+m_s)\,(2\,L_6-L_4)+  m\, (2\,L_8-L_5)\Big]+\cdots \,,
\nonumber\\
&& m_K^2 = B_0\,(m+m_s)+ \frac{16\,B^2_0\,(m+m_s)}{f^2}\,\Big[(2\,m+m_s)\,(2\,L_6-L_4)
\nonumber\\
&& \qquad \qquad \quad\;\; +\, \frac{1}{2}\,( m+m_s)\,(2\,L_8-L_5)\Big]+\cdots\,,
\nonumber\\
&& m_\eta^2 = \frac{2}{3}\,B_0\,(m+2\,m_s) +\frac{128\,B^2_0}{9\,f^2}\,(m-m_s)^2\,(3\,L_7+L_8) 
\nonumber\\
&& \quad\;\;  +\,\frac{32\,B^2_0\,(m+2\,m_s)}{3\,f^2}\,\Big[
(2\, m+m_s)\,(2\,L_6-L_4) +  \frac{1}{3}\, (m + 2\,m_s)\, (2\,L_8-L_5) \Big]+\cdots  \,.
\label{meson-masses-q4}
\end{eqnarray} 
At the one-loop level there are in addition tadpole-type contributions. The terms 
involving the tadpole of the pseudo-scalar mesons were considered already in
\cite{Gasser:1983yg,Gasser:1984gg}. In contrast corresponding structures  involving the tadpole 
with vector mesons are less well studied. Altogether we find
\begin{eqnarray}
&& \Pi^{\rm tadpole}_{P\in [8]}  =  \frac{1}{4\,f^2} \sum_{Q\in [8]} \Big\{-\Big(m_P^2+m_Q^2\Big)\,G_{PQ}^{(T)}
- G_{PQ}^{(\chi)} \Big\}\,\bar I_Q 
\nonumber\\
&& \qquad \quad \;\,+ \, \frac{3}{4\,f^2}\sum_{V\in [9]}
\Big\{- 2\,m_P^2\, G^{(T)}_{PV}    \Big\}\,\bar I_V\,,
\label{tadpole-P}
\end{eqnarray}
with the tadpole function $\bar I_Q$ already recalled in (\ref{def-IQ2}). The vector  meson tadpole $\bar I_V$ follows from $\bar I_Q$ with the replacement $m_Q \to M_V$.
The coefficients $G_{PQ}^{(T)}, G_{PQ}^{(\chi)} $ and $G_{PV}^{(T)} $  are detailed in Tab. \ref{Clebsch-P-Tadpole}. 
The index $P,Q\in[8]$ run over the octet of Goldstone bosons, properly grouped into isospin multiplets. The index $V\in[9]$ runs over the nonet 
of vector mesons. 
While the parameters $B_0$ and $m,m_s$ determine $G_{PQ}^{(T)}, G_{PQ}^{(\chi)} $ the additional vector meson parameters  $g_{1-5}$ 
are probed in $G_{PV}^{(T)} $. 

\begin{table}[t]
\setlength{\tabcolsep}{0.0mm}
\begin{tabular}{l||c|c|c||c|c|c  }
 P  \;\;& \;\;Q  \;\;& $G^{(\chi)}_{PQ}$              & \;\;$G^{(T)}_{PQ}\;\;$      &\;\; V  \;\;  & $G^{(T)}_{PV}$ \\   \hline
        & $\pi$      & $\;(20/3)\,B_0\,m \to   (28/9)\,m_\pi^2  + (8/9)\,m_K^2  -(2/3)\,m_\eta^2  \; $ & $-\frac 83$                 & $\rho$       & $2\,g_1 + 4\,g_2 + \frac{3}{2}\,g_3 - g_5\,$ \\
$\pi$   & $K$        & $(4/3)\,B_0\,(3\,m+m_s) \to (4/3)\,(m_\pi^2 + m_K^2)$ & $-\frac 43$                 & $K^*$        & $2\,g_1 + 2\,g_2 + g_3 - \frac{1}{2} \,g_5\,$ \\ 
	& $\eta$     & $(4/3)\,B_0\,m \to       (2/3)\,m_\pi^2    $ & $0$                         & $\omega$     & $2\,g_1 + \frac{1}{2} \,g_3$ \\
	&            &                                &                             & $\phi$       & $0$ \\ 
\hline 
        & $\pi$      & $B_0\,(3\,m+m_s) \to            m_\pi^2 + m_K^2  $ & $-1$                        & $\rho$       & $\,\frac{3}{2}\,g_1 + \frac{3}{2}\,g_2 + \frac{3}{4}\,g_3 - \frac{3}{8}\,g_5\,$ \\
$K$     & $K$        & $4\,B_0\,(m+m_s) \to            (2/3)\,(3\,m_\eta^2 + 2\,m_K^2 + m_\pi^2) $  & $-2$                        & $K^*$        & $\,3\,g_1 + 3\,g_2 +\frac{3}{2}\,g_3 - \frac{3}{4}\,g_5\,$ \\ 
	& $\eta$     & $(1/3)\,B_0(m+3\,m_s) \to  m_\eta^2 - (1/3)\,m_K^2 $ & $-1$                        & $\omega$     & $\,\frac{1}{2}\,g_1 + \frac{1}{2}\,g_2 +\frac{1}{4}\,g_3 - \frac{1}{8}\,g_5\,$ \\
	&            &                                &                             & $\phi$       & $g_1 + g_2 + \frac{1}{2} \,g_3 - \frac{1}{4}\,g_5\,$ \\ 
\hline 
        & $\pi$      & $4\,B_0m \to                 2\,m_\pi^2 $    & $0$                         & $\rho$       & $\,2\,g_1 +\frac{1}{2}\,g_3$ \\
$\eta$  & $K$        & $(4/3)\,B_0\,(m+3\,m_s)\to 4\,m_\eta^2 - (4\,m_K^2)/3 $& $-4$                        & $K^*$        & $\,\frac{2}{3}\,g_1 + 6\,g_2 +\frac{5}{3}\,g_3 - \frac{3}{2}\,g_5\,$ \\ 
	& $\eta$     & $(4/9)\,B_0\,(m+8\,m_s) \to (2/3)\,(7\,m_\eta^2 - 4\,m_K^2)$ & $0$                         & $\omega$     & $\,\frac{2}{3}\,g_1 +\frac{1}{6}\,g_3$ \\
	&            &                                &                             & $\phi$       & $\frac{8}{3}\,g_1 +\frac{2}{3}\,g_3$ \\
\end{tabular}
 \caption{The coupling constants for vector mesons 
  $G_{PQ}^{(\chi)}, G_{PQ}^{(T)}$ and $G_{PV}^{(T)}$
  with $V\in [9]$ and $P,Q \in [8]$ defined with respect to isospin states.}
\label{Clebsch-P-Tadpole}
\end{table}

At the one-loop level there remain additional contributions involving vertices from ${\mathcal L}_2^{(3)}$ in (\ref{def-L24}), which involve a 
bubble loop integral. We derive their form with
\begin{eqnarray}
&& \Pi^{\rm bubble}_{P\in [8]}  =  \sum_{\substack{Q\in [8],\,V\in [9]}}
\Big(\frac{G^{(P)}_{QV}}{2\,f}\Big)^2 \Bigg\{ -\frac{1}{4}\,\Big(M_V^2-m_Q^2\Big) ^2\,\Delta I_{QV} 
-\frac{1}{4}\,m_P^2 \,\Big(\bar I_Q +\bar I_V \Big)\nonumber\\
&& \qquad \qquad  \qquad \qquad \qquad \qquad \quad \!
-\,\frac{1}{4}\,m_P^2\,\Big( m_P^2 -2\,(m_Q^2+M_V^2)\Big)\,I_{QV}\Bigg\}
\nonumber\\
&& \qquad \quad +\,  \sum_{V,R \,\in [9]}
\Big(\frac{G_{VR}^{(P)}}{2\,f}\Big)^2\Bigg\{  - \frac{1}{4}\, \alpha^{P}_{VR}\,\Big( M_R^2-M_V^2\Big)^2\,\Delta I_{VR}+
\Big(\beta^{P}_{VR}\,m_P^2\, \bar I_V + \beta^{P}_{R\,V}\,m_P^2\,\bar I_R \Big)
\nonumber\\
&& \qquad  \qquad \qquad  \qquad \qquad \quad \;\;- \,\frac{1}{4}\,\alpha^{P}_{VR}\,m_P^2\,\Big( m_P^2-2\,(M_R^2+M_V^2)\Big) \,I_{VR} \Bigg\}\,,
\nonumber\\
&&\alpha_{VR}^{P}=  \frac{(M_V^2+M_R^2 )^2 }{4\,M_V^2\, M_R^2}  \,,\qquad \qquad \qquad \quad 
 \beta_{VR}^{P}  = \frac{ 7\,M_V^4 -10\, M_V^2\, M_R^2 -2\,M_R^4}{32\,M_V^2\,M_R^2} \,,  
\label{loop-P}
\end{eqnarray}
in terms of the scalar bubble functions  $I_{QV}, \Delta I_{QV} $, $I_{VR}, \Delta I_{VR}$ and the previously introduced tadpole integrals $\bar I_Q, \bar I_V, \bar I_R$.  
The loop functions $I_{QV}$ follows from $I_{QR}$ in (\ref{def-bubble}) with the replacements $M_V \to m_P$ together with $M_R \to M_V$. 
In contrast the loop functions $I_{VR}$ follows from $I_{QV}$ with the replacement $m_Q \to M_R$.
For later convenience and in order to avoid any misinterpretation we provide 
the explict representation nevertheless 
\begin{eqnarray}
&& I_{VR} = \bar I_{VR} - \frac{\bar I_R}{2\,M_R^2}-  \frac{\bar I_V}{2\,M_V^2}\,, \qquad \qquad \qquad \qquad \Delta I_{VR} = I_{VR} + \frac{\bar I_V - \bar I_R}{M_V^2- M_R^2}\,,
 \nonumber\\
&& \bar I_{VR} = \frac{1}{16\,\pi^2}
\left\{ 1- \frac{M_V^2-M_R^2}{2\,m_P^2}  \,\log \left( \frac{M_V^2}{M_R^2}\right)
\right.
\nonumber\\
&& \;\quad \;\,+\left.
\frac{p_{V R}}{m_P}\,
\left( \log \left(1-\frac{m_P^2-2\,p_{V R}\,m_P}{M_V^2+M_R^2} \right)
-\log \left(1-\frac{m_P^2+2\,p_{V P}\,m_P}{M_V^2+M_R^2} \right)\right)
\right\} \, ,
\nonumber\\
&& p_{V R}^2 =
\frac{m_P^2}{4}-\frac{M_V^2+M_R^2}{2}+\frac{(M_V^2-M_R^2)^2}{4\,m_P^2}  \,,
\label{def-bubble-VR}
\end{eqnarray} 
where we point at the symmetric definition of the object $\bar I_{VR} = \bar I_{R\,V}$. 
The coefficients $G^{(P)}_{QR}$ and $G^{(P)}_{VR}$ with $P,Q\in [8]$ and $V,\,R\in [9]$ are proportional to the 
coupling constants $h_1$ and $h_2$. They are listed in Tab. \ref{GPQV}.

\begin{table}[t]
 \begin{tabular}{llll}
 $G_{\pi \rho}^{(\pi)} = \sqrt{2}\,h_1 \qquad$       & $G_{KK^{*}}^{(\pi)} = h_1 \qquad $                                  & $G_{\pi K^{*}}^{(K)} = -\frac{\sqrt{3}}{2}\,h_1 \qquad $ & $ G_{K\rho}^{(K)} = \frac{\sqrt{3}}{2}\,h_1 $ \\
 $G_{K\omega}^{(K)} = \frac{1}{2}\,h_1 $             & $G_{\eta K^*}^{(K)} = -\frac{\sqrt{3}}{2}\,h_1 $                          & $G_{K\phi}^{(K)} = -\frac{\sqrt{2}}{2}\,h_1 $            & $ G_{K K^{*}}^{(\eta)} = \sqrt{3}\,h_1$ \\
 \hline
 $G_{\rho\omega}^{(\pi)} = 2\,h_2$                   & $G_{\rho K^*}^{(K)} = \sqrt{3}\,h_2$                                      & $G_{\rho\rho}^{(\eta)} = 2\,h_2$                      & $G_{K^*K^*}^{(\eta)} = -\frac{2}{\sqrt{3}}\,h_2$ \\
 $G_{K^*K^*}^{(\pi)} = 2\,h_2$                 & $G_{\omega K^*}^{(K)} = G_{\phi K^*}^{(K)}/\sqrt{2} =h_2 \quad$           & $G_{\omega\omega}^{(\eta)} = \frac{2}{\sqrt{3}}\,h_2$ & $G_{\phi\phi}^{(\eta)} = -\frac{4}{\sqrt{3}}\,h_2$     \\
 \end{tabular}
  \caption{The coupling constants $G^{(P)}_{QR}$ and $G^{(P)}_{VR}=G^{(P)}_{RV}$ with $P,Q\in [8]$ and $V,\,R\in [9]$. .}
 \label{GPQV}
 \end{table}

A few comments are here in order. All contributions in (\ref{tadpole-P},\ref{loop-P}) comply with their expected power-counting order $Q^4$ in the scenario where 
$m_{P,Q} \sim M_{V,R} \sim Q$. The scale dependence from the loop contributions is balanced by a corresponding dependence of the low-energy constants. In order to establish a strict renormalization 
we have to decompose the low-energy constant $B_0 $ into its power counting moments with
\begin{eqnarray}
 B_0 = \sum_{n=0} M^{2\,n}\,B_0^{(2\,n)}\,.
 \label{decompose-B0}
\end{eqnarray}
This implies the condition
\begin{eqnarray}
&& \mu^2 \frac{\text{d}}{\text{d}\mu^2}  B^{(2)}_0 = - \frac{9\big(4\, g_1 +4\, g_2 +2\, g_3 - g_5\big) }{64 \pi^2 f^2}\,B^{(0)}_0 - \frac{9\, h_1^2 }{256\pi^2 f^2}\,B^{(0)}_0 - \frac{63\, h_2^2 }{256 \pi^2 f^2}\,B^{(0)}_0\,,
\nonumber\\
&& \mu^2 \frac{\text{d}}{\text{d}\mu^2}  \big(2 \,L_6 - L_4 \big) = - \frac{1}{1152 \pi^2} - \frac{3\, h_1^2}{4096 \pi^2}  \,,
\nonumber\\
&& \mu^2 \frac{\text{d}}{\text{d}\mu^2}  \big(3 \,L_7 + L_8 \big) = -\frac{5}{1536\pi^2} \,,
\nonumber\\
&&\mu^2 \frac{\text{d}}{\text{d}\mu^2}  \big(2 \,L_8 - L_5 \big) = +\frac{1}{192\pi^2} - \frac{3 \,h_1^2}{4096\pi^2} + \frac{3 h_2^2}{512\pi^2}\,.
\label{def-Li-running}
\end{eqnarray}

In order to scrutinize the importance of dynamical vector meson degrees of freedom it is useful to match our results to the conventional $\chi$PT expression derived from the flavour SU(3) Lagrangian at the one-loop 
level. This is readily achieved by a further chiral expansion of (\ref{tadpole-P}, \ref{loop-P}) where now the counting rule $m_Q/m_V \sim Q$ has to be applied. 
If truncated to order $Q^4$ the only effect of the vector mesons is a renormalization of Gasser and Leutwyler's 
low-energy constants. With this we find
\allowdisplaybreaks
\begin{eqnarray}
&&B_0^{\rm ren}  = B_0 - \frac{9\, B_0 (4 g_1 +4 g_2 +2 g_3 - g_5)}{64 \pi^2 f^2} \,M^2 \,\log \frac{ M^2}{\mu^2} 
\nonumber\\
&&\qquad \quad
 -\frac{3 \,h_1^2 \,M^2\, B_0 }{512 \,\pi ^2 f^2} \left(1 + 6\, \log \frac{M^2}{\mu^2}\right) 
 -  \frac{3 \,h_2^2 \,M^2\, B_0 }{256 \,\pi ^2 f^2}\left(16+21 \,\log \frac{M^2}{\mu^2}\right) \,,
\nonumber\\ \nonumber\\
&& 2\,L_6^{\rm ren}-L_4^{\rm ren}  =  2\,L_6-L_4  - \frac{h_1^2}{8192 \,\pi ^2}\left(1+6 \, \log \frac{M^2}{\mu ^2}\right)\,, \qquad
\nonumber\\
&& 2\,L_8^{\rm ren}-L_5^{\rm ren}  =  2\,L_8-L_5 - \frac{h_1^2}{8192 \,\pi ^2} \left(-7+6\, \log \frac{M^2}{\mu ^2}\right)
 - \frac{h_2^2}{4096 \,\pi ^2} \left(-40-24\log \frac{M^2}{\mu ^2}\right) \,,
\nonumber\\
&&  3\,L_7^{\rm ren}+L_8^{\rm ren}  =  3\,L_7+L_8\,. \phantom{\Bigg)}
\label{fullrenormPQ2Q4}
\end{eqnarray} 
The contributions proportional to $h^2_1$ have been considered in the literature before \cite{Meissner:1987ge,Birse:1996hd,Bijnens:1999sh,Bruns:2004tj,Terschlusen:2016cfw,Terschlusen:2016kje}. Our results are consistent with the recent 
study \cite{Terschlusen:2016cfw,Terschlusen:2016kje}. The effect of the coupling constant $h_2$ 
is typically not considered in resonance saturation approaches to Gasser and Leutwyler's low-energy constants \cite{Meissner:1987ge,Ecker:1988te,Birse:1996hd,Bijnens:1999sh}. In particular its contribution to $B_0$ and 
$2\,L_8-L_5$ is sizable, a factor 10-30 larger than the corresponding terms proportional to $h_1^2$. 
Again it is convenient to explore the size of the loop effects at the particular renormalization scale $\mu = M$. In this case we obtain
\begin{eqnarray}
 B_0^{\rm ren}/B_0 = 1 -  \underbrace{ 3\, \Big( h_1^2/32+ h_2^2\Big)}_{\simeq \,16.9}\,\left( \frac{M}{4\,\pi \,f } \right)^2\,,
\end{eqnarray}
a huge correction term primarily caused by the $h_2 $ term.  
In contrast to our findings in the vector meson sector we observe a significant size of the loop correction that poses 
a challenge to the dimensional counting rules.

Note however that the result (\ref{estimate-Mchi}) is not unexpected 
since the typical ratio 
\begin{eqnarray}
\frac{M}{\Lambda_\chi}= \frac{M}{4\pi\, f} \sim 1 \,,
\end{eqnarray}
is probed in (\ref{fullrenormPQ2Q4}), which is of order one numerically in any case. While for sufficiently large values of $N_c$ we have $\Lambda_\chi \geq  \Lambda_{\rm HG}$ by assumption this is not the case 
for the physical choice with $N_c = 3$  and $\Lambda_\chi \simeq 1$ GeV.  In turn all terms proportional to $(M/\Lambda_\chi)^n$ need to be summed in our approach. This is the target of 
the following development.

We wish to identify renormalized low-energy parameters $l^{\rm ren }$ which have a decomposition of the following form
\begin{eqnarray}
 l^{\rm ren } = \sum_{n=0}^\infty  l_n \,\left(\frac{M}{4\pi f} \right)^{2\,n} \,\qquad {\rm with} \qquad l^{\rm ren } \sim \Lambda^{\rm dim [l^{\rm ren }]}_{HG}\,,
 \label{def-resummation}
\end{eqnarray}
and therefore justify the application of the dimensional counting rules. While for sufficiently large $N_c$ the renormalized coupling constants can be conveniently matched 
to the parameters of the hadrogenesis Lagrangian in perturbation theory, at $N_c =3$ a suitable summation scheme is required. Such a scheme is readily devised by exploiting 
the simple observation: the particular combination
\begin{eqnarray}
 \left(\frac{M}{4\pi f} \right)^2 \frac{B_0\,m_{\rm quark}}{ M^2}  \sim Q^2 \,,
 \label{def-part}
\end{eqnarray}
is consistent with the  dimensional counting rule the  hadrogenesis Lagrangian is based on. Note that the second factor in (\ref{def-part}) arises naturally if a loop contribution involving vector mesons 
is expanded in powers of the quark masses. We conclude that if we absorb any terms proportional to powers of the ratio $M/\Lambda_\chi$ into 
the low-energy parameters of the chiral Lagrangian then necessarily the particular combination (\ref{def-part}) arises. For instance in the chiral domain  it followed for the accordingly renormalized loop 
contribution
\begin{eqnarray}
 \Pi^{\rm bubble}_P \Big|_{\rm renormalized }\sim B^2_0\,m^2_{\rm quark}\,,
\end{eqnarray}
for sufficiently small quark masses. All non-perturbative effects in  $M/\Lambda_\chi$ are moved into the renormalized low-energy parameters. The important observation is that there is no need to actually 
perform the infinite summation explicitly. Since such a summation should be performed in accordance with the symmetries of the hadrogenesis Lagrangian the generic structure of the result must resemble 
the generic structure of a perturbative computation at $\Lambda_\chi \geq \Lambda_{\rm HG}$. Thus it suffices to express the bare coupling constants in terms of the renormalized coupling constants order by order in 
perturbation theory. Technically it is more economical to devise a suitable subtraction scheme for the loop functions involving vector mesons \cite{Lutz2000,Semke2005,Lutz:2014oxa}. If performed at the level of the 
Passarino Veltman functions such a renormalization scheme is symmetry conserving not violating any chiral Ward identities.

We introduce the following subtraction rules
\begin{eqnarray}
&& \bar I_{V} \to 0 \,, \qquad \qquad  \qquad \qquad \quad \qquad\quad  \bar I_{R} \to 0 \,, \qquad \qquad
\nonumber\\
&& I_{QV} \to \bar I_{QV} +  \frac{1}{4}\,\frac{1}{(4\, \pi)^2}\,,\qquad \qquad \quad \;\;I_{VR} \to  \bar I_{VR} + \frac{1}{(4\, \pi)^2}\,, 
 \label{def-replacement-P}
\end{eqnarray}
where all low-energy parameters within the renormalization scheme (\ref{def-replacement-P}) receive an upper index $'{\rm r\,}'$ as to discriminate them from the 
bare parameters used initially. Indeed 
in (\ref{tadpole-P}) and (\ref{loop-P}) all contributions from the vector meson loops to the low-energy parameter $B^r_0$  vanish identically. Note that the scale invariant bubble structures $\Delta I_{QV}$ and 
$\Delta I_{VR}$ remain untouched. Given the subtraction scheme (\ref{def-replacement-P}) it is feasible to use physical meson masses everywhere without picking up an uncontrolled dependence on the 
renormalization scale $\mu$. Like for the vector meson polarization tensor it suffices to reinterpret the appropriate counter term contributions 
\begin{eqnarray}
&& m_\pi^2 =2\,B^{\rm r}_0\,m  - \frac{8\,m_\pi^2}{f^2}\, \Bigg[ m_\pi^2\,(L^{\rm r}_5 - 2\,L^{\rm r}_8) + (2\,m_K^2 + m_\pi^2)\,(L^{\rm r}_4 - 2\,L^{\rm r}_6) \Bigg] + \cdots \,,
\nonumber\\
&& m_K^2 = B^{\rm r}_0\,(m + m_s) - \frac{8\,m_K^2}{f^2}\,\Bigg[ m_K^2\,(L^{\rm r}_5 - 2\,L^{\rm r}_8) + \frac{3}{2}\,(m_\eta^2 + m_\pi^2)\,(L^{\rm r}_4 - 2\,L^{\rm r}_6) \Bigg]+ \cdots \,,
\nonumber\\
&& m_\eta^2 = \frac{2}{3}\,B^{\rm r}_0\,(m+2\,m_s) - \frac{8\,m_\eta^2}{f^2}\,\Bigg[ m_\eta^2\,(L^{\rm r}_5 - 2\,L_8) - 3\,(m_\eta^2 - 2\,m_K^2)\,(L^{\rm r}_4 - 2\,L^{\rm r}_6) \Bigg]
\nonumber\\
&& \qquad \qquad \quad\;\; +\, \frac{128}{40\,f^2}\,\Bigg[ 13\,m_\eta^4 - 8\,m_\eta^2\,m_K^2 - 8\,m_K^4 + 3\,m_\pi^4 \Bigg]\,(L^{\rm r}_8 +3\,L^{\rm r}_7)+ \cdots\,.
\label{meson-masses-q4-rewrite}
\end{eqnarray} 
in terms of physical masses. Note that this result was derived already in \cite{Guo:2018a}.  
We assure that with (\ref{meson-masses-q4-rewrite}) and the replacements provided in Tab. \ref{Clebsch-P-Tadpole} scale invariant results for the pion, kaon and eta meson masses 
are obtained.

The subtraction scheme implies in particular that there remains no explicit dependence on any of the unknown low-energy parameters $g_n$. 
Moreover  a modification for the 
expression for the renormalized low-energy constants with
\begin{eqnarray}
&& \mu^2 \frac{\text{d}}{\text{d}\mu^2}  \big(2 \,L^{\rm r}_6 - L^{\rm r}_4 \big) = - \frac{1}{1152 \pi^2}
 - \frac{h_1^2}{4096 \,\pi ^2} \,, \qquad 
 \mu^2 \frac{\text{d}}{\text{d}\mu^2}  \big(3 \,L^{\rm r}_7 + L^{\rm r}_8 \big) = -\frac{5}{1536\pi^2} \,,
\nonumber\\
&&\mu^2 \frac{\text{d}}{\text{d}\mu^2}  \big(2 \,L^{\rm r}_8 - L^{\rm r}_5 \big) = +\frac{1}{192\pi^2} 
- \frac{3h_1^2}{4096 \,\pi ^2}\,,
\label{res-DeltaLi}
\end{eqnarray}
is observed. We note that the large contribution proportional to $h_2^2$ in $2\,L^{\rm r}_8 - L^{\rm r}_5$ is reduced by a factor $2/5$ as compared to the original expression (\ref{fullrenormPQ2Q4}).  
Suppose we have determined our low-energy parameters $L^{\rm r}_i$ from some data set. How would we confront them with the conventional low-energy parameters of Gasser and Leutwyler, which we 
denote here by $\bar L_i$? The required relations are provided with
\begin{eqnarray}
&& 2\,\bar L_6-\bar L_4  =  2\,L^{\rm r}_6-L^{\rm r}_4  - \frac{h_1^2}{4096 \,\pi ^2} \log \frac{M_\chi^2}{\mu^2}\,, \qquad \qquad 3\,\bar L_7+\bar L_8  =  3\,L^{\rm r}_7+L^{\rm r}_8\,,
\nonumber\\
&& 2\,\bar L_8-\bar L_5  =  2\,L^{\rm r}_8-L^{\rm r}_5 
- \frac{h_1^2}{8192 \,\pi ^2} \bigg( -7 + 6\log \frac{M_\chi^2}{\mu^2} \bigg) + \frac{h_2^2}{256 \,\pi ^2} \,.
\end{eqnarray}
We affirm that the scale dependence of the parameters, $\bar L_i$, resembles the one of the 
conventional $\chi$PT approach without dynamical vector mesons, i.e. the formulae in (\ref{def-Li-running}) taken at $h_i=0$.

\begin{figure}
\centering
\includegraphics[width=0.7\textwidth]{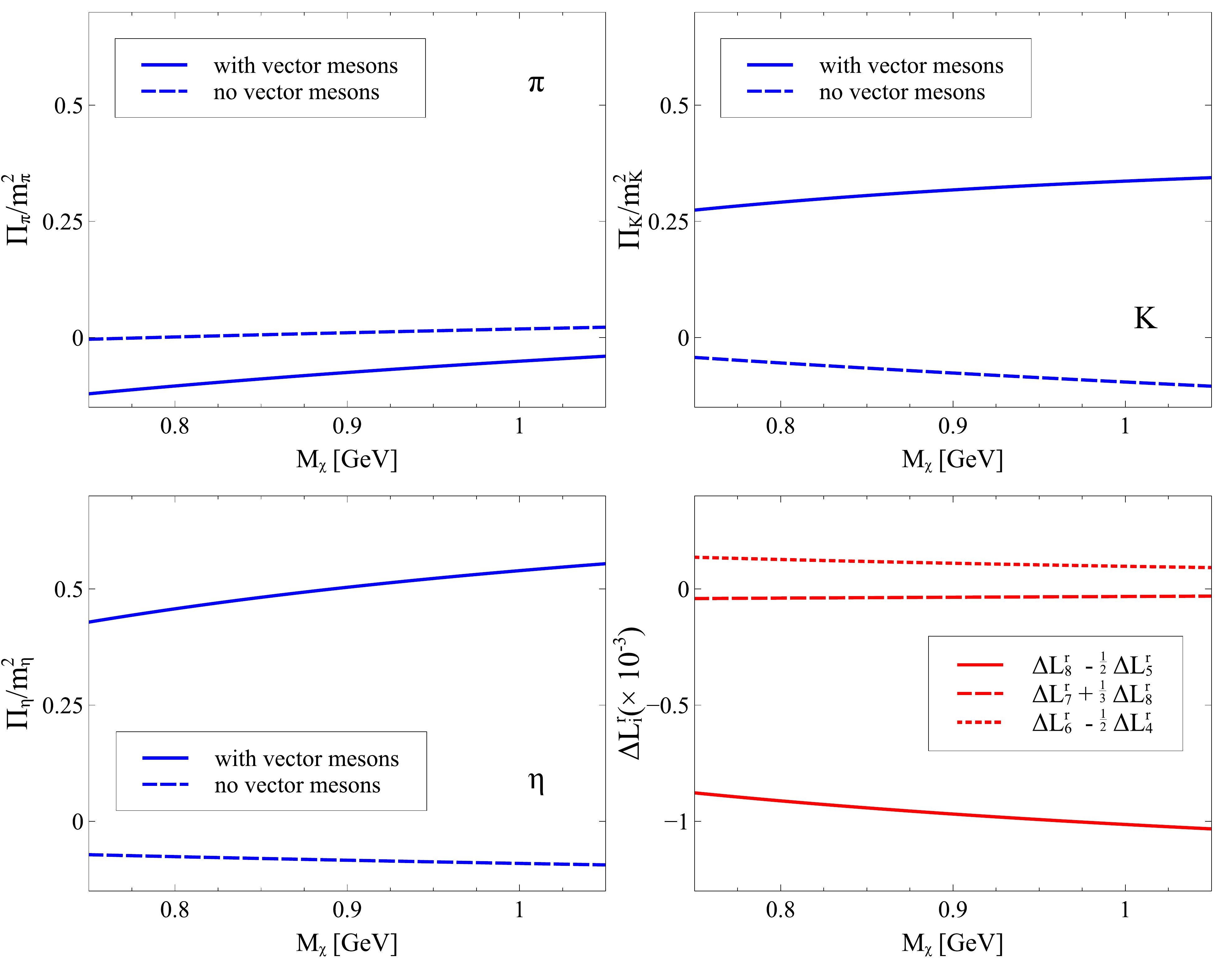}
\caption{
The pseudo-scalar meson polarization  $\Pi_P/m_P^2$ are plotted as a function of $M_\chi$, at $\mu=M_\chi$. For the meson masses inside the loop functions leading order expressions are used as described in the text with 
$\epsilon_\omega = \epsilon_\phi = 0$ and $L^{\rm r}_i = 0$. While the solid lines include the effect of vector-meson loop contributions,  the dashed lines leave the latter contributions out.  
In the last plot  specific $ \Delta L^{\rm r}_i$ are shown as functions of $M_\chi$. The $L^{\rm r}_i =\Delta L^{\rm r}_i $ are determined such that their effect would move the solid lines back ontop of the dashed lines.}
\end{figure}

\begin{figure}
\centering
\includegraphics[width=0.7\textwidth]{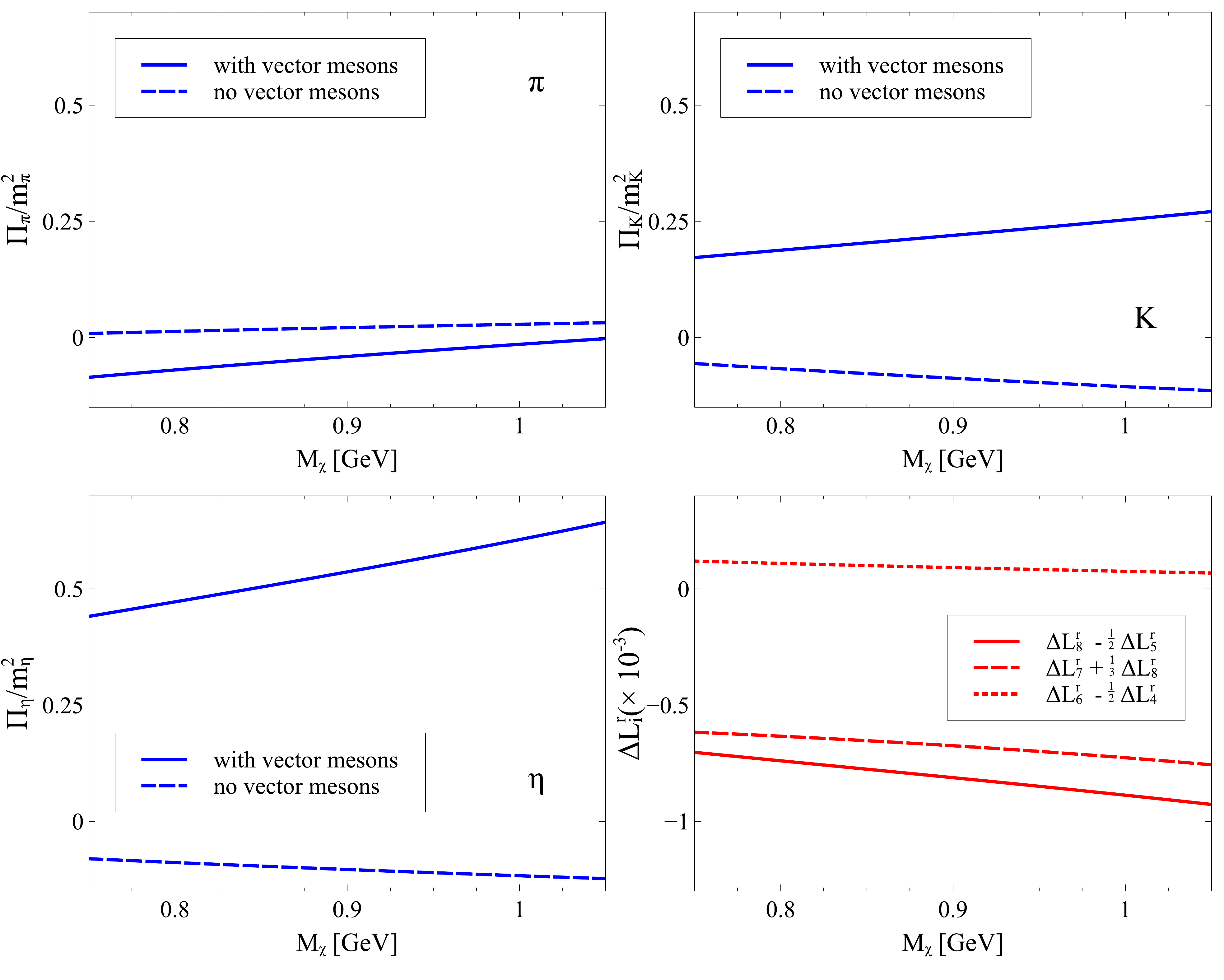}
\caption{The pseudo-scalar meson polarization  $\Pi_P/m_P^2$ are plotted as a function of $M_\chi$, at $\mu=M_\chi$ and $L^{\rm r}_i = 0$. For the meson masses inside the loop functions physical values are assumed. 
For the mixing angles we use $\epsilon_\omega = 0.45$ and  $\epsilon_\phi = 0.058$. While the solid lines include the effect of vector-meson loop contributions,  the dashed lines leave the latter contributions out.  
In the last plot  specific $ \Delta L^{\rm r}_i$ are shown as functions of $M_\chi$. The $L^{\rm r}_i =\Delta L^{\rm r}_i $ are determined such that their effect would move the solid lines back ontop of the dashed lines.}
\end{figure}

It is instructive to compare our result to the well established one-loop expression of $\chi$PT in the absence of dynamical vector mesons. The corresponding expressions for the pion, kaon and 
eta meson masses can readily be 
recognized in (\ref{meson-masses-q4}) and (\ref{tadpole-P}). 
We illustrate  the role of the dynamical vector mesons in the ratio $\Pi^{}_P/m_P^2$  as a function of $M_\chi$ at $\mu = M_\chi$. For this purpose 
we determine the product of $B^{\rm r}_0$ and the quark masses from the physical pion and kaon mass 
\begin{eqnarray}
 2\,B_0^{\rm r} \, m \simeq m_\pi^2\,, \qquad \qquad \qquad m_s \simeq  27.3 \,m\,,
 \label{def-scenario}
\end{eqnarray}
in terms of the Gell-Mann Oakes Renner relation for the pion mass and the latest quark mass ratio from the PDG \cite{Patrignani:2016xqp}. 
Since we are after the typical size of loop effects the contributions from the renormalized tree-level parameters are switched off with  
$L^{\rm r}_i = 0 $. Like for our vector-meson mass study we consider two cases both using $f^{\rm r} = 90$ MeV. In Fig. 5 we show  the ratios  $\Pi^{}_P/m_P^2$ as determined from 
(\ref{tadpole-P}, \ref{loop-P}) with $M_{V,R} \to M_\chi $ and the  kaon and eta meson masses approximated by the Gell-Mann Oakes Renner relations, i.e. 
$m_K^2=B_0^{\rm r} \, (m + m_s) $ and 
$m_\eta^2 = \frac{1}{3}(4\,m_K^2 - m_\pi^2)$. 
The Fig. 6 shows the same ratios 
evaluated with physical values for the masses of the pion, kaon and eta meson as well as all vector mesons. In both figures the subtraction rules (\ref{def-replacement-P}) are imposed.
Two lines are shown for the pion, kaon and eta meson ratios always. While the solid lines show the effect including the contributions of the vector mesons the dashed lines 
follow with $h_1 = h_2 = h_3= 0$ strictly for which there are no contributions form vector mesons. 

In all cases we find a significant effect from the vector-meson loop contributions. It is pointed out that such effects cannot be simply absorbed into the low-energy 
constants $L^{\rm r}_i$ as was worked out with (\ref{res-DeltaLi}). At the particular choice $\mu = M_\chi$  the vector meson loop contributions renormalize exclusively the 
particular combination
\begin{eqnarray}
  2\,\bar L_8-\bar L_5  =  2\,L^{\rm r}_8-L^{\rm r}_5 + \underbrace{\frac{7\,h_1^2}{8192 \,\pi ^2} + \frac{h_2^2}{256 \,\pi ^2}}_{\simeq \,3.2\times 10^{-3}}\,,
\end{eqnarray}
for which we provide its numerical estimate. With this one may have expected  $\Delta L^{\rm r}_7 + \Delta L^{\rm r}_8/3 =
\Delta L^{\rm r}_6 - \Delta L^{\rm r}_4/2 = 0$ and $\Delta L^{\rm r}_8- \Delta L^{\rm r}_5 /2\simeq - 1.6$ in Fig. 4 or Fig. 5. 
The latter values are far away from the results presented in the figures. 
We conclude that there are significant non-linear structures from the vector meson loops that 
must not be expanded in the quark masses as suggested by conventional $\chi$PT.

\clearpage 

\subsection{Decay constants of the Goldstone bosons at the one-loop level}

We close this work with a study of the one-loop contributions to the decay constants $f_P$ of a Goldstone bosons of type $P$. 
According to  \cite{Gasser:1984gg} the conventional approach leads to the following expressions
\begin{eqnarray}
&&f^{\chi-{\rm PT}}_\pi  =  f-\frac{1}{f}\,\bar{I}_\pi - \frac{1}{2\,f}\,\bar{I}_K + \underbrace{\frac{8\,B_0\,m}{f}}_{\to \,4\,m_\pi^2/f}\,L_5 
+ \underbrace{\frac{8\,B_0\,(2\,m + m_s)}{f}}_{\to \,4\,(2\,m_K^2+m_\pi^2)/f}\,L_4 \,,
\nonumber\\
&&f^{\chi-{\rm PT}}_K  =  f-\frac{3}{8\,f}\,\bar{I}_\pi - \frac{3}{4\,f}\,\bar{I}_K - \frac{3}{8\,f}\bar{I}_\eta +\underbrace{ \frac{4\,B_0\,(m + m_s)}{f}}_{\to \,4\,m_K^2/f}\,L_5 
+ \underbrace{\frac{8\,B_0\,(2\,m + m_s)}{f}}_{\to\, 6\,(m_\eta^2+ m_\pi^2)/f}\,L_4\,,
\nonumber\\
&& f^{\chi-{\rm PT}}_\eta  =  f-\frac{3}{2\,f}\,\bar{I}_K + \underbrace{\frac{8\,B_0\,(m + 2m_s)}{3\,f}}_{4\,m_\eta^2/f}\,L_5 
+ \underbrace{\frac{8\,B_0\,(2\,m + m_s)}{f}}_{\to\,12\,(2\,m_K^2-m_\eta^2)/f}\,L_4\,,
\label{decay-tree}
\end{eqnarray}
with the tadpole integrals as given in (\ref{tadpole-P}). 
Before providing the additional contributions that arise from the presence of dynamical vector mesons we further illuminate our scheme formulated 
in terms of physical masses. Within the conventional $\chi$PT approach the pion, kaon and $\eta$ meson masses that enter the tadpole integrals $\bar I_Q$ in (\ref{decay-tree}) need 
to be approximated by the leading order expressions, i.e. $m_\pi^2\to 2\,m\,B_0$ etc. If done so the expressions for the decay constants will not depend on the 
renormalization scale $\mu$. However, it would clearly be instrumental if we could keep the physical masses inside the loops without giving up on the rigor of conventional $\chi$PT. 
An initial attempt where one simply kept the tadpole terms with physical masses suffers from an uncontrolled scale dependence of the resulting expressions for the decay constants. 
Is it possible to identify the higher order terms that would again lead to scale invariance? Such terms should be determined by a renormalization group equation.
Indeed it is possible to construct such terms unambiguously. All what is needed is to reinterpret the quark mass terms in (\ref{decay-tree}) by suitable combinations of the pion, kaon and $\eta$ meson masses
as indicated by the replacement rules in (\ref{decay-tree}). We assure that with the later the physical masses in the tadpole terms can be used without being punished by a scale dependence in the 
decay constants. 

We turn now to the contributions from dynamical vector meson degree of freedom. 
Like for  the vector meson masses such terms will renormalize the chiral limit 
value of $f_P$ away from the parameter $f$. Altogether, for the decay constants we find
\begin{eqnarray}
&& f_{P\in[8]} = f^{\chi-\rm{PT}}_P + \frac{3}{4\,f}\sum_{V\in [9]}\, G^{(T)}_{PV} \,\bar I_V + \Bigg( \frac{f}{2}\,\frac{\partial}{\partial \,m_P^2}- \frac{f}{m_P^2}\,\Bigg)\,\Pi^{\rm bubble}_P \,,
\label{loop-fP}
\end{eqnarray}
where we point at the close correspondence of (\ref{loop-P}) and (\ref{loop-fP}). In particular all coefficient $G^{(T)}_{PV}$
have been introduced before in (\ref{loop-P}) and are listed in Tab. \ref{Clebsch-P-Tadpole}.  

Like in the previous section we first determine the scale dependence of the relevant low-energy parameters in strict perturbation theory. For this purpose we need to decompose $f$ into its power counting moments with
\begin{eqnarray}
 f = \sum_{n=0}^\infty M^{2\,n}\,f^{(2\,n)}\,.
 \label{decompose-f}
\end{eqnarray}
While the leading order moment $f^{(0)}$ remains scale invariant the second order moment $f^{(2)}$ does depend on the renormalization scale. 
Altogether we derive 
\begin{eqnarray}
&& \mu^2\, \frac{\text{d}}{\text{d}\mu^2} f^{(2)} =  +\frac{ 9\, (4 g_1 +4 g_2 +2 g_3 - g_5)}{128\, \pi^2 f^{(0)}} 
+\, \frac{18\, h_1^2}{1024 \pi^2 f^{(0)}} + \frac{63\, h_2^2}{512 \pi^2 f^{(0)}} \,,
\nonumber\\
&& \mu^2\, \frac{\text{d}}{\text{d}\mu^2} L_4 = - \frac{1}{256 \pi^2} + \frac{3 h_1^2}{4096 \pi^2}\,,  \qquad \qquad
 \mu^2 \,\frac{\text{d}}{\text{d}\mu^2} L_5 = -\frac{3}{256\pi^2} + \frac{9 h_1^2}{4096\pi^2}\, .
\end{eqnarray}
It remains to identify the renormalized low-energy parameters. Again they follow upon a quark-mass expansion of the loop function that involve the dynamical vector mesons. We 
introduce
\begin{eqnarray}
&& f^{\rm ren} = f + \frac{ 9\, (4 g_1 +4 g_2 +2 g_3 - g_5)}{128\, \pi^2 f^{\rm}}\, M^2 \,\log \frac{ M^2}{\mu^2}
\nonumber\\
&& \quad \quad + \,\frac{3\, h_1^2\, M^2}{1024 \pi^2 f^{\rm }} \,\bigg( 1 +6 \log \frac{ M^2}{\mu^2} \bigg)
	+ \frac{3\, h_2^2\, M^2}{512 \pi^2 f^{\rm }}\,\bigg( 16 +21 \log \frac{ M^2}{\mu^2} \bigg)\,,
\nonumber\\
&& L_4^{\rm ren} = L_4 +\frac{h_1^2}{8192\pi^2}\,\bigg(1 +6 \log \frac{ M^2}{\mu^2} \bigg)\,,    \qquad \quad 
 L_5^{\rm ren} = L_5 +\frac{3\,h_1^2}{8192 \pi^2}\,\bigg( 1 +6 \log \frac{ M^2}{\mu^2} \bigg)\, .
 \label{ren-f}
\end{eqnarray}
Like we observed for the low-energy parameters $B_0$ and $2\,L_7-L_5$ there is a significant contribution from the $h_2$ coupling constant in the renormalized expression for the 
low-energy parameter $f$. The latter is about a factor 20 larger than the corresponding term induced by the $h_1$ coupling constant. The results for the renormalized $L_4$ and $L_5$ 
parameters are in line with expressions given previously in the literature \cite{Terschlusen:2016cfw,Terschlusen:2016kje}. There is no contribution from $h_2$ in this case. We iterate that 
it is mandatory to resum all terms proportional to $(M/\Lambda_\chi)^n$. The expressions (\ref{ren-f}) as they stand are not significant.

\begin{figure}
\centering
\includegraphics[width=0.7\textwidth]{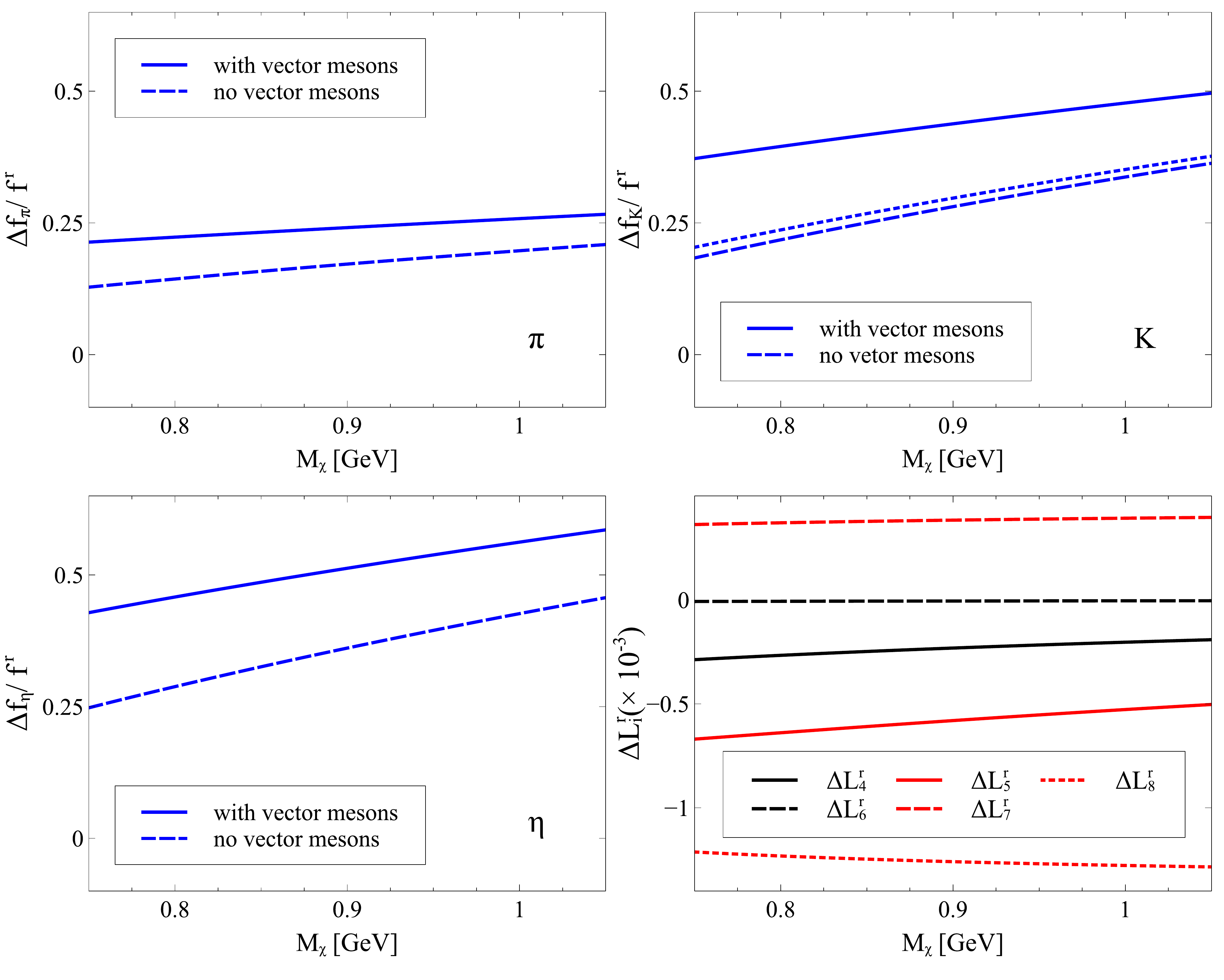}
\caption{The ratios $\Delta f_P/f^{\rm r}$ are plotted as a function of $M_\chi$ at $\mu=M_\chi$ and $L^{\rm r}_i =0$. For the meson masses inside the loop functions leading order expressions are used as in Fig. 5. 
While the solid lines include the effect of vector-meson loop contributions,  the dashed lines leave the latter contributions out.  
In the last plot  specific $ \Delta L^{\rm r}_i$ are shown as functions of $M_\chi$. The $L^{\rm r}_i =\Delta L^{\rm r}_i $ are determined such that their effect would move 
the solid lines in the pion and eta meson box  of Fig. 5 and Fig. 7 back on top of the dashed lines. The dotted line in the kaon box of Fig. 7 shows the effect of the $\Delta L^{\rm r}_i$ on the kaon decay constant.}
\end{figure}

Again we impose the subtraction rules 
(\ref{def-replacement-P}) in (\ref{loop-fP}) which are expected to generate the desired summation effects (\ref{def-resummation}) we are after. 
We assure the reader that as an immediate consequence of (\ref{def-replacement-P}) the low-energy parameter $f^r$ is not renormalized by loop effects. 
This implies
\begin{eqnarray}
\Big[ f_P - f_P^{\chi-PT} \Big]_{renormalized}  \sim \frac{B^{\rm r}_0\,m_{\rm quark}}{ (4\pi\, f^{\rm r})^2} \,f^{\rm r} \,, 
\end{eqnarray}
in the chiral domain with the quark masses approaching the chiral limit. We note that there is no explicit dependence left on any of the unknown low-energy parameters $g_n$. 
Moreover, we can safely use physical masses in all loop expression. Scale invariant results arise for the decay constants if and only if the replacement rules indicated already in (\ref{decay-tree}) 
are imposed. It remains to identify the low-energy parameters $\bar L_4$ and $\bar L_5$ for which we obtain
\begin{eqnarray}
	&& \bar L_4 = L^{\rm r}_4 + \frac{h_1^2}{4096\pi^2} \log \frac{M_\chi^2}{\mu^2}\, ,
	\qquad \qquad \bar L_5 = L^{\rm r}_5 + \frac{3\, h_1^2}{4096\pi^2} \log \frac{M_\chi^2}{\mu^2}\, .
\end{eqnarray}

\begin{figure}
\centering
\includegraphics[width=0.7\textwidth]{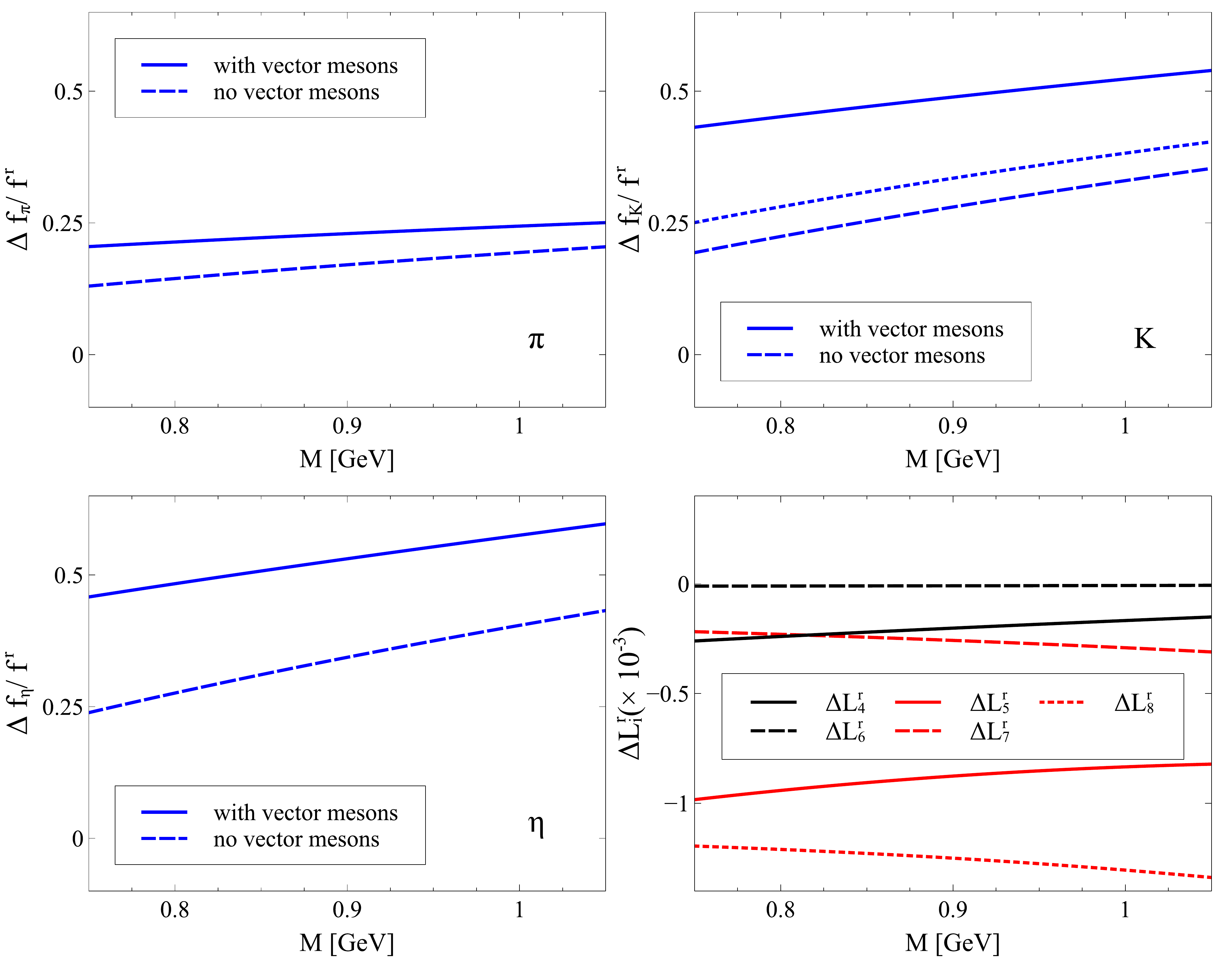}
\caption{The ratios $\Delta f_P/f^{\rm r}$ are plotted as a function of $M_\chi$ at $\mu=M_\chi$ and $L^{\rm r}_i =0$. For the meson masses inside the loop functions physical are used as in Fig. 6. 
While the solid lines include the effect of vector-meson loop contributions,  the dashed lines leave the latter contributions out.  
In the last plot  specific $ \Delta L^{\rm r}_i$ are shown as functions of $M_\chi$. The $L^{\rm r}_i =\Delta L^{\rm r}_i $ are determined such that their effect would move 
the solid lines in the pion and eta meson box  of Fig. 6 and Fig. 8 back on top of the dashed lines. The dotted line in the kaon box of Fig. 8 shows the effect of the $\Delta L^{\rm r}_i$ 
on the kaon decay constant.}
\end{figure}

We are now prepared to illustrate the role of vector meson loop contributions in the decay constants of the Goldstone bosons.  In Fig. 7 and Fig. 8 we plot the normalized ratio $f_P/f^{\rm r} - 1$ as a function 
of $M_\chi$ at $\mu = M_\chi$ together with
\begin{eqnarray}
f^{\rm r} = 90\, {\rm MeV}\,,\qquad \qquad  L_4^{\rm r } = L_5^{\rm r } = 0\,. 
\end{eqnarray}
Like in the previous Fig. 5 and Fig. 6  we show the results of using approximated and physical meson masses respectively.  
Two lines are shown for the normalized ratios of the pion, kaon and eta meson decay constants. While the solid lines show the effect including the contributions of the vector mesons the dashed lines 
follow with $h_1 = h_2 = h_3= 0$ strictly for which there are no contributions from vector mesons. The low-energy parameters $L^{\rm r}_4= \Delta L^{\rm r}_4$ and $L^{\rm r}_5= \Delta L^{\rm r}_5$ 
can be adjusted to cancel the effect of the vector meson loop contributions to the pion and eta meson decay constants. Given the scenario (\ref{def-scenario}) with $f^{\rm r} = 90$ MeV we determine the 
low-energy constants as a function of $M_\chi$. We observe again the the use of physical meson masses in the loop functions play an important role. 

We recall that if the vector-meson loop contributions would be approximated well by conventional $\chi$PT structures at order $Q^4$ we would have 
obtained the specific values
\begin{eqnarray}
\Delta L^{\rm r }_4 = \Delta L^{\rm r }_5 = \Delta L^{\rm r }_6 = 0 \,,\qquad \quad \Delta L^{\rm r }_8 \simeq -1.6 \,,\qquad \quad \Delta L^{\rm r }_7 \simeq 0.5\,.
\end{eqnarray}
As is clearly shown by Fig. 7 and Fig. 8 we are far from such a situation. Thus we conclude it is important to consider dynamical vector meson degrees of freedom in an chiral extrapolation 
attempt of meson masses in QCD.

\clearpage

\section{Summary and outlook}

In this work we scrutinized the hadrogenesis Lagrangian, a chiral $SU(3)$ interaction with explicit vector meson degrees of freedom in the tensor field representation. 
Based on the leading order interaction the one-loop contributions to the vector meson masses was computed in application of dimensional counting rules. We found that 6 parameters from the original 
version of the Lagrangian need to be moved to higher order as to arrive at a consistent renormalization program. This is an important finding since this further 
increases the predictive power of the hadrogenesis Lagrangian. 

The subtle interplay of the hadrogenesis mass gap scale $\Lambda_{\rm HG}$ and the chiral symmetry breaking scale $\Lambda_\chi$ 
was discussed. The dimensional counting rules rely on the assumption $M/\Lambda_{\rm HG}$, with $M$ the vector meson mass in the chiral and large-$N_c$ limit of QCD. 
At sufficiently large $N_c$ with $\Lambda_\chi \geq \Lambda_{\rm HG}$ the hadrogenesis Lagrangian can be applied in perturbation theory. 
For the physical choice $N_c=3$ with $\Lambda_\chi < \Lambda_{\rm HG}$ a partial summation of all terms proportional to $M/\Lambda_\chi \sim 1 $ is required as to arrive at significant results. 
It was suggested that this can be achieved by a suitable renormalization scheme. First numerical estimates for the size of the one-loop corrections for the vector meson masses were provided given such a framework.
The results are in line with the expectation of the dimensional counting rules.

The work was supplemented by computations of the one-loop corrections of the masses and decay constants of the Goldstone bosons. 
The size of the loop contributions for vector meson degrees of freedom was illustrated by a series of figures, which suggest good convergence properties of the effective field theory.

Further steps that are required to consolidate our findings on the crucial importance of dynamical vector meson degrees of freedom. The result obtained in this work can be used for an attempt 
to describe the quark-mass dependence of unquenched QCD lattice simulation data  
on the vector mesons as well as on the masses and decay constants of the the Goldstone bosons. Such data are expected to determine some of the so far unknown low-energy constants of the 
hadrogenesis Lagrangian. Additional constraints on the form of the hadrogenesis Lagrangian are expected from a one-loop study of the vector meson scattering amplitudes. 
It is also left to investigate the role of the $\eta$' meson, which was not considered in the current study.

 \clearpage

\section{Appendix}

The chiral expansions of scalar bubbles read
\begin{eqnarray}
	&& I_{QP} = \frac{1}{16 \pi ^2}\bigg[ 1-\log \frac{ M^2  }{\mu ^2} + \frac{m_P^2}{ M^2} \bigg(1-\log \frac{ m_P^2}{ M^2}\bigg) + \frac{m_Q^2}{ M^2} \bigg(1-\log \frac{ m_Q^2}{ M^2}\bigg)\nonumber\\
	&& \qquad \qquad -\frac{m_Q^4}{2 M^4} -\frac{m_P^4}{2 M^4} - \frac{m_Q^2 m_P^2}{M^4} \bigg(\log \frac{m_P^2}{\bar M^2}+\log\frac{  m_Q^2}{ M^2}\bigg) + \dots \bigg] \\
	&& I_{QR}   = \frac{1}{16\pi^2} \bigg[ 1 - \log \frac{ M^2}{\mu^2} - \frac{m_Q\, \pi}{ M} 
	+\frac{m_Q^2 }{2 M^2} \bigg(2 -\log \frac{ m_Q^2}{ M^2}\bigg) +\frac{m_Q^3}{8 M^3}  -\frac{m_Q^4}{12 M^4} + \dots \bigg]. \\
	&& I_{RT} = \frac{1}{16\pi^2} \bigg[  1-\frac{\pi }{\sqrt{3}} - \log \bigg(\frac{ M^2}{\mu^2} \bigg) +\dots \bigg],
\end{eqnarray}
where the vector-meson masses are assigned to be their chiral limit $M_V = M_R = M_T = M$.

The strict chiral expansion of the scalar bubbles,
\begin{eqnarray}
&& \bar I_{QV} =\frac{1}{16 \pi^2}\bigg[ \frac{m_P^2}{2\, M^2} + \frac{\,m_Q^2}{ M^2} \log\bigg( \frac{m_Q^2}{ M^2} \bigg) + \frac{m_P^4 + 9\, m_P^2\, m_Q^2}{6 M^4} + \frac{m_Q^2}{M^4}\big( m_P^2 + m_Q^2 \big) \log\bigg( \frac{m_Q^2}{ M^2} \bigg) + \dots \bigg].\nonumber\\
&& \bar I_{VR} =\underbrace{-\frac{1}{16 \pi^2}}_{\text{from the chiral limit of $I_{VR}(0)$, so $\Delta I_{VR}$ is free from this term}}  + \frac{1}{16 \pi^2} \bigg( \frac{m_P^2}{6\,M^2} + \frac{m_P^4}{60\, M^4} \bigg) +\dots
\end{eqnarray}
In the above expansion, the vector meson mass are evaluated at the chiral limit $M_V = M_R = 
M$.

\clearpage

\bibliography{literature}
\bibliographystyle{apsrev4-1}
\end{document}